\documentclass[twocolumn,aps,prx,showpacs,amsmath,superscriptaddress,longbibliography,notitlepage]{revtex4-1}
\usepackage{amssymb}
\usepackage{mathrsfs}
\usepackage{graphicx}
\usepackage{float}
\usepackage[caption=false]{subfig}
\usepackage[pdftex,colorlinks,linkcolor={red!75!black},citecolor={red!75!black},urlcolor={blue!75!black}]{hyperref}
\usepackage{verbatim}
\usepackage{epstopdf}
\usepackage[normalem]{ulem}
\usepackage{xcolor}
\usepackage{bm}
\usepackage{soul}
\usepackage{ulem}
\renewcommand{\Re}{\operatorname{Re}}
\renewcommand{\Im}{\operatorname{Im}}
\newcommand{\clr}{\color{red!75!black}}

\newcommand{\Rnum}[1]{\uppercase\expandafter{\romannumeral #1\relax}}

\def\LLH{\textcolor{cyan}}

\begin{document}
\title{Geometry-dependent skin effect and anisotropic Bloch oscillations in a non-Hermitian optical lattice}
\author{Yi Qin}
\affiliation{Guangdong Provincial Key Laboratory of Quantum Metrology and Sensing $\&$ School of Physics and Astronomy, Sun Yat-Sen University (Zhuhai Campus), Zhuhai 519082, China}
\author{Kai Zhang}\email{phykai@umich.edu}
\affiliation{Department of Physics, University of Michigan Ann Arbor, Ann Arbor, Michigan, 48105, United States}
\author{Linhu Li}\email{lilh56@mail.sysu.edu.cn}
\affiliation{Guangdong Provincial Key Laboratory of Quantum Metrology and Sensing $\&$ School of Physics and Astronomy, Sun Yat-Sen University (Zhuhai Campus), Zhuhai 519082, China}

\begin{abstract}
	The interplay between the non-Hermiticity and dimensionality gives rise to exotic characteristics in higher dimensions, with one representative phenomenon known as the geometry-dependent skin effect (GDSE), which refers to that the localization of extensive eigenstates depends on the system's geometry under open boundary conditions. 
	In this paper, we demonstrate the emergence of GDSE in a two-dimensional $sp$ optical ladder lattice with on-site atom loss, which can be manifested by anisotropic dynamics of Bloch oscillations in the bulk of the system.
	By applying a static force in different directions, the wave-packet dynamics retrieve the complex energy spectra with either nonzero or zero spectral winding number, indicating the presence or absence of skin accumulation in the corresponding directions, respectively. 
	Our results reveal that the GDSE has an intrinsic anisotropic bulk dynamics independent of boundary conditions, and offer its realization and detection in quantum systems. 	
\end{abstract}

\maketitle
\emph{{\clr Introduction}.---}~Ultracold quantum gases in optical lattices provide an ideal platform for investigating intriguing quantum physics on both single-particle and many-body levels~\cite{bloch2005ultracold,Bloch2008,bloch2008many,giorgini2008theory,Sun2011,goldman2016topological,zhang2018topological}.
Rich lattice structures with highly controllable parameters
can be engineered in these systems with state-of-art quantum engineering techniques,
such as double-well lattices constructed by multi-frequency optical potentials~\cite{Sebby2006,Anderlini2007,Trotzky2008,Wirth2010,Soltan2011,Soltan22011,Jo2012,Tarruell2012,Zhou2011,CaiZi2011}, 
and nontrivial topological bands induced by spin-orbital coupling~\cite{Sato2009,Jiang2011,Lutchyn2010}, artifical gauge field~\cite{Goldman2010}, or inter-orbital couplings~\cite{Lixiao2013}.
While generally considered as closed quantum systems,
ultracold quantum gases are also capable of demonstrating novel non-Hermitian physics describing open quantum systems, where non-Hermiticity can be effectively introduced by laser-induced particle dissipation~\cite{Li2019,ren2022chiral}. 
In particular, realization of the seminal non-Hermitian skin effect (NHSE)~\cite{Yao2018,Martinez2018}, 
which refers to accumulation of a majority of eigenstates to the system's boundary,
has been theoretically predicted and experimentally implemented with ultracold atoms in either real space or momentum space~\cite{li2020topological,liang2022dynamic,guo2022theoretical,li2022BEC,zhou2022engineering}.

Being a boundary phenomenon, boundary conditions play an important role in determining the behavior of NHSE~\cite{
Koch2022,Budich2020,Li2021,GuoCX2021_PRL,Liu2021,Mu2022}.
In higher dimensional systems, richer boundary conditions of different geometries and dimensions lead to more sophisticated variations of NHSE, e.g. disclination and dislocation NHSE induced by defect boundaries~\cite{sun2021geometric,bhargava2021non,panigrahi2022non,schindler2021dislocation,Kai2022DDS,Ashvin2019_PRL}, and hybrid skin-topological effect with eigenstate accumulation at lower-dimensional boundaries~\cite{li2020topological,Lee2019hybrid,zou2021observation,li2022gain,zhu2022hybrid,ou2023nonH}.
Recently, a special type of geometry-dependent skin effect (GDSE) has been discovered in two and higher dimensional lattices~\cite{Kai2022NC}, where skin accumulation of eigenstates always occur under the open boundary conditions (OBCs) of certain geometries, representing the universality of NHSE beyond one-dimensional systems~\cite{HuiJiang2022,Murakami2022,HYWang2022}. 
To date, implementations of GDSE have been proposed and experimentally realized in various systems including atomic arrays~\cite{Wang2022}, photonic crystals~\cite{Fang2022}, acoustic crystals~\cite{Zhou2023,Wan2023}, and active mechanical lattices~\cite{wang2023experimental}, yet an explicit setup realizing GDSE in quantum systems still remains elusive. 

In this paper, we propose a realization of GDSE in a two dimensional (2D) optical lattice with $sp$ orbital coupling induced by a double-well potential.
NHSE is forbidden along either $x$ or $y$ direction of a square lattice by the system's symmetry, 
but emerges along oblique directions in a diamond lattice in the presence of orbital-dependent atom loss.
The skin accumulation of eigenstates are found to have opposite directions for different oblique momenta, generating a reciprocal GDSE toward one-dimensional (1D) boundaries of a 2D diamond lattice.
We then further unveil that the GDSE leads to anisotropic Bloch oscillations for driving forces of different orientations, 
which can serve as a probe of skin accumulation along corresponding directions.
Remarkably, Bloch oscillations involve only bulk dynamics and are insensitive to boundary conditions, thus provide an efficient scheme for detecting different types of NHSE in optical lattices, where sharp open boundaries with desired geometries are usually difficult to be implemented.

\begin{figure}[t]
	\begin{center}
		\includegraphics[width=1\linewidth]{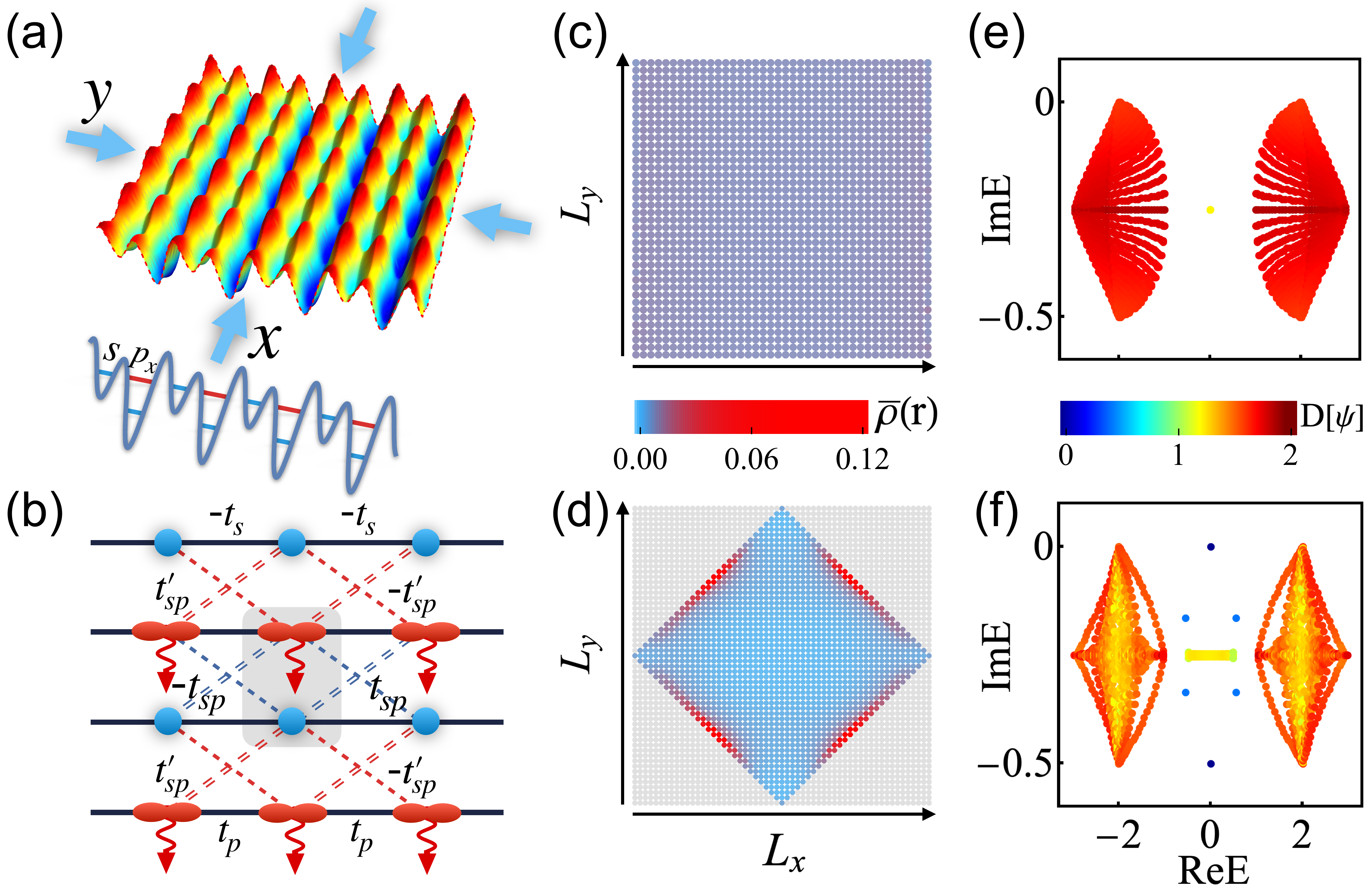}
		\par\end{center}\protect\caption{\label{fig:1} 
		\textbf{The non-Hermitian two-dimensional $sp$ ladder model.}  (a) A 2D optical lattice with double-well structure along $y$ direction. (b) Schematic illustration of the 2D non-Hermitian $sp$ ladders, with the curved arrows indicating the loss on $p_x$ orbital. (c)(d) The average density of all eigenstates in square and diamond  geometries with $\gamma=0.5$. The corresponding system size are $L_{x}\times L_{y}=40\times 40$ and  $L_{x}\times L_{y}=57\times 57$, respectively. (e)(f) The spectra of the fully OBC system with square and diamond geometries, respectively, with different colors indicating the fractal dimension $D[\psi]$ of each eigenstate. In (f), extra 1D edge states emerge in the band gap with $D[\psi_{\rm edge}]\approx1$. Other parameters are $ t_{s}=1,t_{p}=1,t_{sp}=1,t_{sp}^{\prime}=0.5$. }
\end{figure}

\emph{{\clr Two-dimensional non-Hermitian $sp$ ladder model}.---}~We consider a system of noninteracting Fermi gas in a 2D optical potential
$V(x,y) = {V_x}{\sin ^2}\frac{2\pi}{\lambda_L}x + {V_1}{\sin ^2}\frac{2\pi}{\lambda_L}y + {V_2}{\sin ^2}(\frac{4\pi}{\lambda_L}y + \phi/2)$, with $V_x$ ($V_1,V_2$) describing the strength of a static single-well (double-well) potential along $x$ ($y$) direction, and $\phi$ is a phase shift of the double-well potential, as shown in Fig.~\ref{fig:1}(a).
With properly designed optical potential, the $s$ and $p_x$ orbitals of the double wells respectively can have roughly the same energy, generating a $sp$ ladder lattice~\cite{Lixiao2013}, as illustrated in Fig. \ref{fig:1}(b).
The orbital-dependent particle loss can be induced with a resonant optical beam transferring atoms to an excited state, which introduces non-Hermiticity to the system~\cite{Li2019}. 

Based on this setup, the tight-binding Hamiltonian of the 2D lattice can be expressed as
\begin{equation}\label{MT_RealHam}
	H=\sum_{\bm{r}} c^{\dagger}_{\bm{r}} T_x c_{\bm{r}+ x} \pm 
	c^{\dagger}_{\bm{r}} T_{xy} c_{\bm{r}\pm x+y}  + h.c.
	-i\sum_{\bm{r}}\gamma c^\dagger_{\bm{r},p_x}c_{\bm{r},p_x},
\end{equation}
where $c^{\dagger}_{\bm{r}}=(c^{\dagger}_{\bm{r},s},c^{\dagger}_{\bm{r},p_x})$ 
represents a pair of creation operators of Fermions for $s$ and $p_x$ orbital of the double wells, $\bm{r}$ indicates the lattice sites.
$T_x = \begin{pmatrix} - t_s & - t_{sp} \\ t_{sp} & t_p
\end{pmatrix}$ 
and 
$T_{xy}  = \begin{pmatrix} 0 & 0 \\ t^{\prime}_{sp} & 0
\end{pmatrix}$ 
are the intra-cell and inter-cell hopping matrices, respectively.
The derivation of the tight-binding parameters from lattice potential is detailed in Supplemental Materials~\cite{SupMat}.
The corresponding Bloch Hamiltonian of the 2D $sp$ lattice reads
\begin{equation}
  \mathcal{H}({k_x},{k_y}) = \boldsymbol{h}(k_x,k_y) \cdot \boldsymbol{\sigma}-i\gamma(\sigma_0-\sigma_z)/2 ,
\end{equation}
where $\bm{\sigma}=(\sigma_0,\sigma_x,\sigma_y,\sigma_z)$ are the $2\times2$ identity matrix and Pauli matrices,
and $\bm{h}(k_x,k_y)$ is a four-component vector describing the Hermitian part of the Hamiltonian, with ${h_0}(k_x,k_y) = ({t_p} - {t_s})\cos {k_x}$, 
${h_x}({k_x},{k_y}) =  - 2t'_{sp}\sin {k_y}\sin {k_x}$, 
$h_y(k_x,k_y) = 2 t_{sp} \sin{k_x} + 2 t^{\prime}_{sp} \cos{k_y} \sin{k_x}$, ${h_z}(k_x,k_y) = -(t_p + t_s)\cos{k_x}$.

To describe the localization of OBC eigenstates and unveil the possible NHSE of the system, we define the average density of all eigenstates as 
\begin{equation}\label{eq:saden}
	\bar{\rho}(\bm{r})=\sum_{\alpha,n}\rho_n(\alpha,\bm{r})/N,
\end{equation}
where $\rho_n(\alpha,\bm{r})=|\psi_{n,\alpha,\bm{r}}|^2$, 
$\psi_{n,\alpha,\bm{r}}$ represents the wave amplitude of the $n$-th eigenstate $\psi_n$ at the $\alpha$ orbital of the lattice site $\bm{r}$, and $N$ is the total number of lattice sites. 
We display $\bar{\rho}(\bm{r})$ for the OBC system with square and diamond geometries in Fig.~\ref{fig:1}(c) and (d), respectively. 
It is seen that the eigenstates are localized to the boundaries on the diamond geometry but extended over bulk on the square geometry, indicating the existence of GDSE. 
To further characterize the localization dimension of eigenstates, we illustrate in Fig.~\ref{fig:1}(e) and (f) the fractional dimension (FD) for each eigenstate, defined as~\cite{Mac2019,Backer2019} 
\begin{equation}
   D[\psi] =  - \ln \left[ {\sum_{\alpha=s,p}\sum_{\bm{r}} {{{\left| {{\psi _{\alpha,\bm{r}}}} \right|}^4}} } \right]/\ln \sqrt N.
\end{equation}
By definition, a state $\psi$ uniformly distributed in the 2D bulk, 1D edges, and a single corner shall have $D[\psi]=2,1,0$ respectively. 
As shown in Fig.~\ref{fig:1}(e), almost all eigenstates in bulk bands have $D[\psi]\approx2$ for square geometry. 
In contrast, for diamond geometry in Fig.~\ref{fig:1}(f), we observe $D[\psi]\gtrsim1$ for the majority of eigenstates in bulk bands, whose number is proportional to the volume of the lattice~\cite{SupMat}, indicating the GDSE in our system. 

\begin{figure}
	\begin{center}
		\includegraphics[width=.9\linewidth]{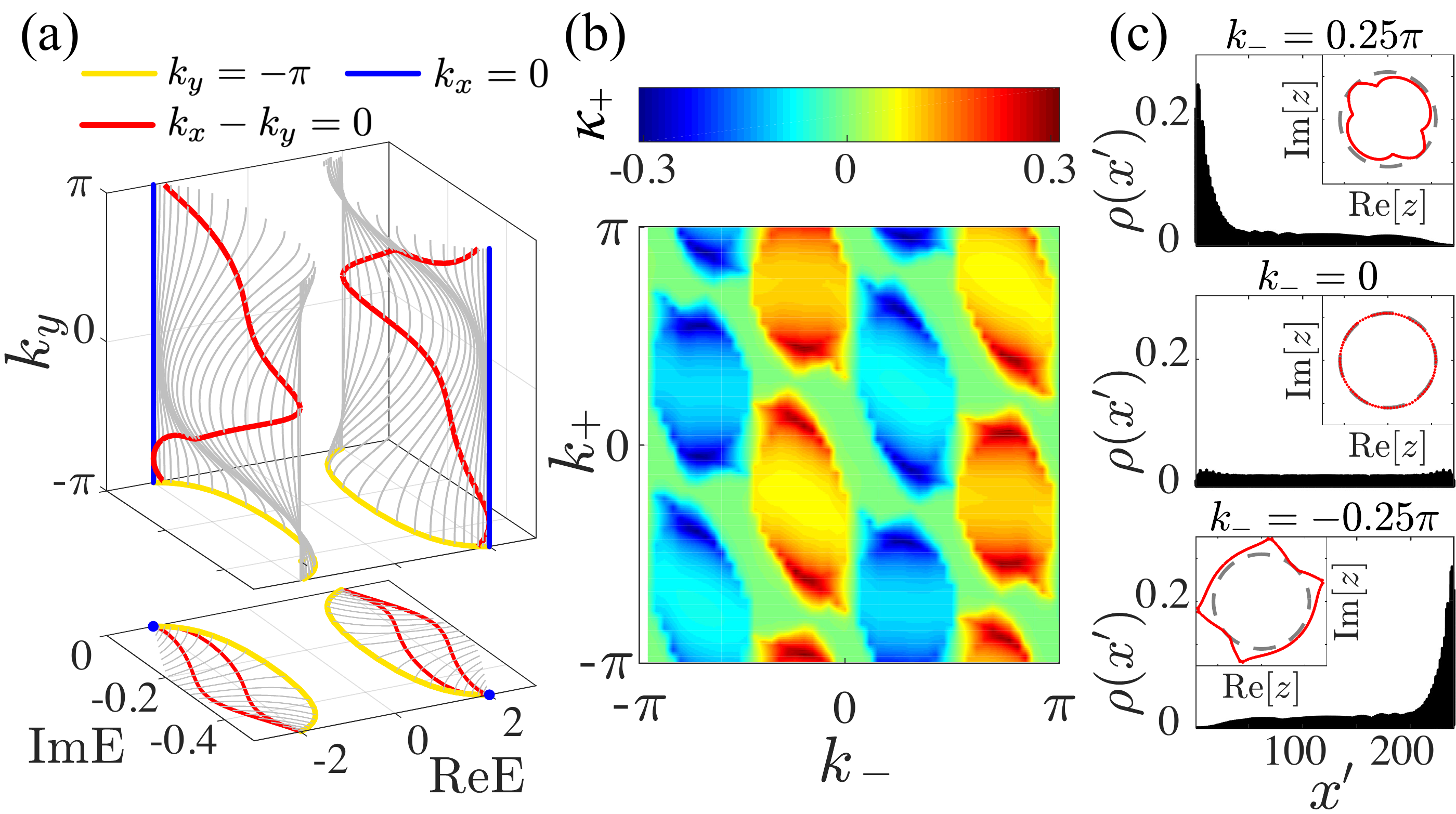}
		\par\end{center}
	\protect\caption{\label{fig:2} 
		\textbf{ Bloch spectral winding and 1D GBZ of the 2D $sp$ ladder model.}
		(a) The upper panel is double PBC spectra in $E-k_y$ space,  and the blue, yellow, and red trajectories represent spectra with $k_x=0$, $k_y=-\pi$, $k_x-k_y=0$, respectively. 
		The lower panel shows the projection of these trajectories on the complex energy plane. 
		(b) The inverse localization length $\kappa_+$ of skin modes with open boundaries along the $(x+y)$ direction. Different signs of $\kappa_+$ correspond to opposite localization directions. 
		(c) The spatial density of eigenstates with OBC along $(x+y)$ direction and several fixed $k_-$. 
		The corresponding 1D GBZs with $z:=e^{i(k_++i\kappa_+)}$, compared to BZ (dashed gray unit circle), are shown by the red curves in the inset. 
		} 
\end{figure}

\emph{{\clr Symmetry-protected GDSE}.---}~A preliminary understanding of the GDSE can be obtained by analyzing the symmetries of the system Hamiltonian. 
The Hamiltonian respects a mirror and a transpose-mirror symmetries along $x$ and $y$ directions, represented by
\begin{eqnarray}
\sigma_z\mathcal{H}(k_x,k_y)\sigma_z = \mathcal{H}(-k_x,k_y),\label{mirror}\\
\sigma_z \mathcal{H}^T(k_x,k_y)\sigma_z  = \mathcal{H}(k_x,-k_y),\label{t_mirror}
\end{eqnarray}
respectively.
As a consequence, the Bloch spectral winding number must vanish along these two directions, which ensures the absence of NHSE for the system under square geometry~\cite{Kai2020,Okuma2020_PRL,borgnia2020nonH}.
As exemplified in Fig.~\ref{fig:2}(a), the PBC spectrum forms arcs (points) enclosing no area for a 1D trajectory along $k_y=-\pi$ ($k_x=0$) in the complex energy plane, and loops with nontrivial spectral winding along an oblique direction of $k_+=k_x+k_y$ ($k_-=k_x-k_y$) for fixed $k_-$ ($k_+$), which corresponds to the absence and emergence of NHSE in square and diamond geometries, respectively. 

To characterize the skin effect in our system, we calculate the 1D generalized Brillouin zone (GBZ)~\cite{Yao2018,yokomizo2019non} corresponding to accumulation of eigenstates with open boundaries along a given direction~\cite{SupMat}. 
Specifically, the GBZ with OBC in $x+y$ direction is given by a complex analytic continuation of the crystal momentum $k_+\rightarrow k_+ - i\kappa_+(k_+,k_-)$, where $H(k_+ - i\kappa_+,k_-)$ reproduces the $(x+y)$-OBC spectrum (i.e., a stripe geometry along $x-y$ direction), and $\kappa_+$ represents the inverse localization length for each skin mode~\cite{PhysRevB.99.201103}. 
Similarly, another inverse localization length $\kappa_-(k_+,k_-)$ can be defined to describe skin modes under $(x-y)$-OBC.
Note that the transpose-mirror symmetry in Eq.~\eqref{t_mirror} becomes $\sigma_z \mathcal{H}^T(k_+,k_-)\sigma_z  = \mathcal{H}(k_-,k_+)$, 
ensuring that the skin modes toward $x+y$ and $x-y$ directions have the same inverse localization lengths for symmetry-related momenta, i.e. $\kappa_+(k_+,k_-)=\kappa_-(k_-,k_+)$. 
Although $\kappa_\pm$ may not directly predict the GDSE and OBC spectrum for the system with a diamond geometry,
they describe the net strength of skin localization along one oblique direction when that of the other direction is balanced by PBC. 
As shown in Fig.~\ref{fig:2}(b), $\kappa_+$ take different positive and negative values for different crystal momenta $k_-$, reflecting a reciprocal type of skin accumulation toward opposite $(x+y)$ and $-(x+y)$ directions respectively. 
Several typical examples of the 1D GBZ and skin modes with fixed $k_-$ are further displayed in Fig.~\ref{fig:2}(c). 
It shows that the corresponding eigenstates experience left and right skin accumulation for GBZ falling inside and outside the BZ respectively, and are extended in the bulk when the GBZ and BZ overlap. 

 \begin{figure}
	\begin{center}
		\includegraphics[width=1\linewidth]{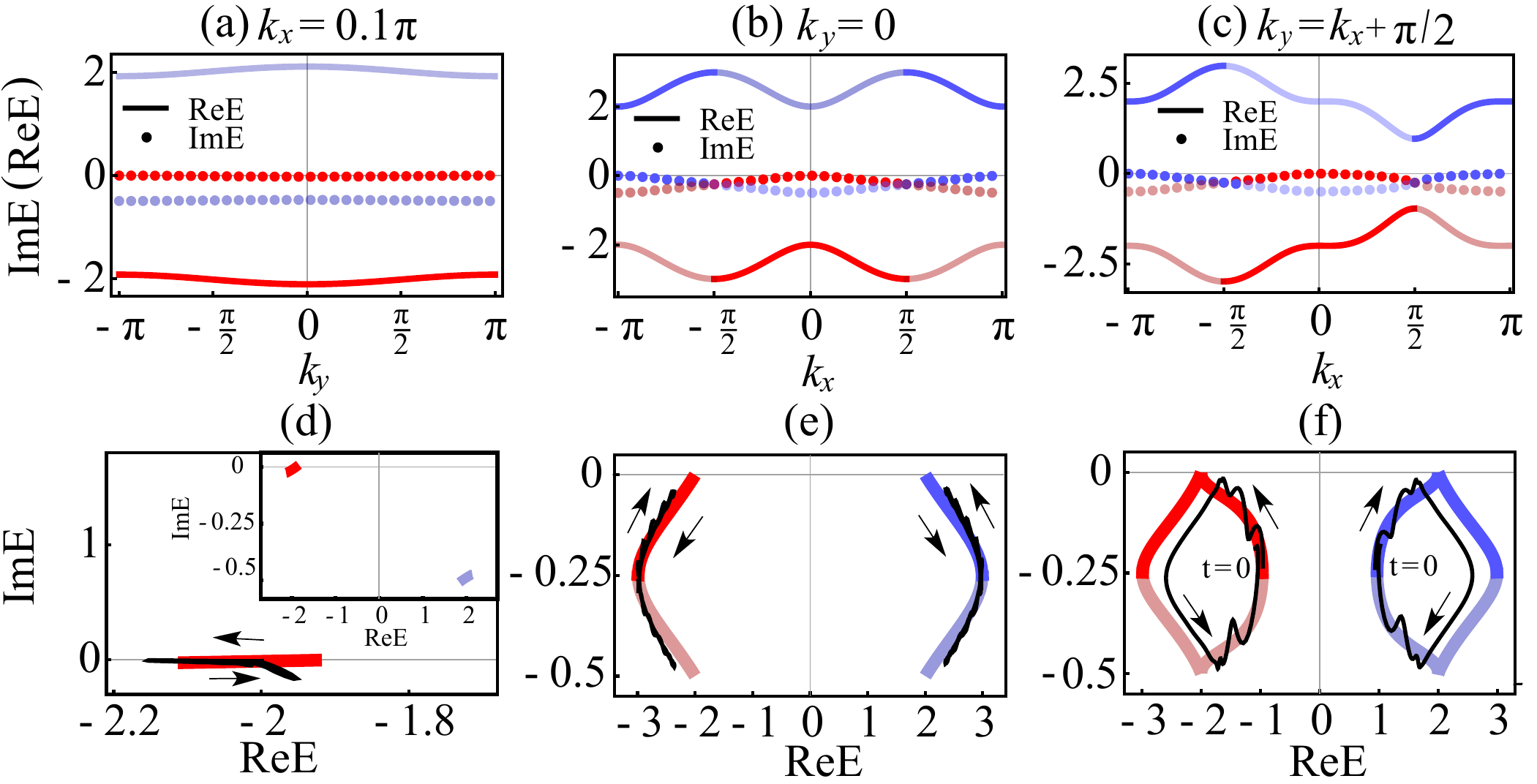}
		\par\end{center}
	\protect\caption{\label{fig:3} 
		\textbf{Complex spectra from the wave-packet dynamics under static forces in different directions.} 
		(a)-(c) The real (lines) and imaginary (dots) band structures with $k_x=0.1\pi$, $k_y=0$, and $k_-=-\pi/2$, respectively. 
		Eigenvalues with larger (lower) imaginary part have greater (lesser) opacity. 
		(d)-(f) The complex spectra corresponding to band structures with the same colors in (a)-(c), respectively,
		and energy trajectories obtained from the wave-packet dynamics (black curves).
		(d) displays only one of the two bands, whose full spectrum is shown in the inset.
		The initial wave packet are chosen to have $\bm{k}_0=(0.1\pi,0)$ in (d), $(\pi/2,0)$ in (e), and $(\pi/2,\pi)$ in (f), 
		with their corresponding static forces being $(F_x, F_y)=(0.25, 0)$, $\pm(0, 0.25)$, and $\pm(0.25, 0.25)$ respectively. 
		Plus (minus) sign reflects the evolution toward positive (negative) $k_x$ directions, reconstructing the energy band with blue (red)
		color.
		In each of these cases, $u_{\bm{k}_0}$ for the initial wave-packet is chosen to be the eigenstate of target energy band of $\mathcal{H}(\bm{k}_0)$. 
		Other parameters are the same as in Fig.\ref{fig:1}. } 
\end{figure}

\begin{figure*}
	\begin{center}
	\includegraphics[width=1\linewidth]{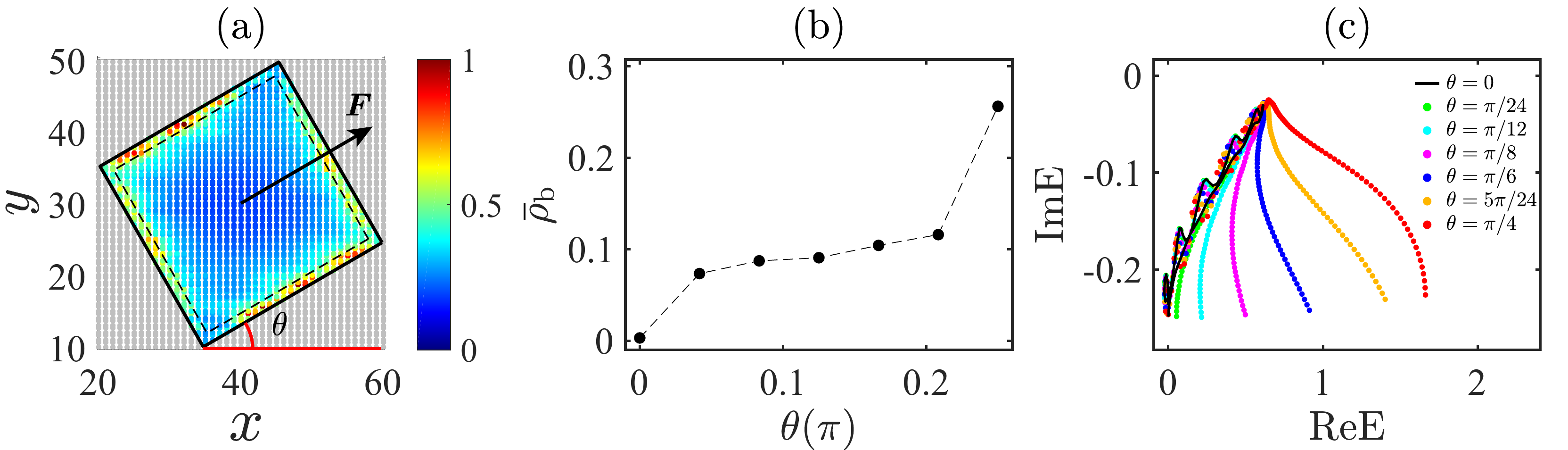}
		\par\end{center}
	\protect\caption{\label{figRotate} \textbf{NHSE on different open-boundary geometries and the corresponding retrieved energy spectra.} 
	\textbf{(a)} The density distribution with $\theta=\pi/6$. 
	$\theta$ indicates the angle between the red line and red dashed line. 
	\textbf{(b)} The weight of $\bar{\rho}(\bm{r})$ in Eq.(\ref{fig:3}) on the specified boundary section versus rotation angle $\theta$. 
	\textbf{(c)} Energy spectra retrieved from the wave-packet dynamics with static force ${\bm F}=(F_x,F_y)$ [black arrow in (a)], where $F_x=F\cos\theta$ and $F_y=F\sin\theta$. 
	${\bm k}_0$ for the initial is chosen to be $(-\pi/2,0)$, and we consider only the dynamics for the time interval $t\in(0,t_{\rm fin})$, where $F_x t_{\rm fin}=\pi$.
	The values of $\theta$ are chosen to be the same as in (b).
	Other parameters are the same as Fig.~\ref{fig:3}. } 
\end{figure*} 

\emph{{\clr Anisotropic bulk dynamics in Bloch oscillations}.---}~Due to the reciprocity of response in GDSE~\cite{Kai2022NC}, methods used to probe non-reciprocal NHSE, such as directional signal amplification~\cite{WenTanXue2021} or edge accumulation of particle populations under long-time dynamics~\cite{XiaoLei2020NP}, are no longer applicable to GDSE in our model. 
Here, we adopt Bloch oscillations to detect the GDSE by showing their anisotropic bulk dynamics. 
We consider the dynamics of an initial wave packet
$$|\psi_{0} \rangle=A\exp( - ( \bm{r}-\bm{r}_0)^2/\sigma_0+i \bm{k}_0 \bm{r} ) \, |u_{\bm{k}_0}\rangle$$ 
in response to a static force $ \bm{F}=F_x\hat{x}+F_y\hat{y}$,
with $\bm{k}_0$ the center momentum and $A$ the normalization factor, and $|u_{\bm{k}_0}\rangle$ represents an eigenvector of Hamiltonian $\mathcal{H}(\bm{k}_0)$. 
The wave-packet width $\sigma_0=4.5$ is chosen in the remaining part of this paper. 
For a non-Hermitian system, its complex spectrum $E(\bm{k})$ is related to the non-unitary Bloch oscillations by means of the semi-classical equations \cite{Longhi2009,Longhi2015,Graefe2016,GongPRX2018}
\begin{equation}\label{eq:BO_E}
	\frac{d \langle\bm{r}\rangle_t}{dt} = \nabla_{\bm{k}}\Re{E(\bm{k})}; \quad
	\frac{d\ln{\mathcal{N}_t}}{dt} = 2 \Im{E(\bm{k})},
\end{equation}
where $\langle\bm{r}\rangle_t=\langle {\psi (t)} | \bm{r}|{\psi (t)} \rangle$ represents the center of mass of the wave packet at time $t$, 
the total probability ${\cal N}_t=\left\langle {\psi (t)} | {\psi (t)} \right\rangle$ is generally less than $1$ in a dissipative system, and $\bm{k}=\bm{k}_0-\bm{F}t$.
As $t$ increases, the momentum $\bm{k}$ will start from $\bm{k}_0$ and run through BZ in the direction of $\bm{F}$ with period $T$, 
with $T=2\pi/F_{x/y}$ for $F_{y/x}=0$, and $T=2\pi/{\rm gcrd}(F_x,F_y)$ for both nonzero $F_x$ and $F_y$, 
where ${\rm gcrd}(a,b)$ represents the greatest common rational divisor of $a,b\in \mathbb{Q}^+$. 
Therefore, the complex spectrum can be reproduced from the wave-packet dynamics in a period $T$, which unveils the information of NHSE in the direction of $\bm{F}$, as numerically verified by several examples in Supplemental Materials~\cite{SupMat}.  

Following the above discussion, the GDSE in our model can be identified by examining wave-packet dynamics by imposing $\bm{F}$ in different directions,
which will unveil the spectral properties and corresponding NHSE along these directions.
Yet for such a non-Hermitian multi-band system, it is crucial to pay attention to the validity of the adiabatic theorem,
since the dynamics shall be dominated by the energy band with larger imaginary energy at each $\bm{k}$,
and a state may jump between bands provided their imaginary energy spectrum cross each other, leading to the breakdown of the adiabatic theorem~\cite{Silberstein2020}.
The instantaneous state at the jump is sensitive to the loss rate and other parameters, and thus the dynamical behavior afterward becomes unpredictable.
In Fig.~\ref{fig:3}(a) to (c), we show real and imaginary parts of band structures along three typical directions with fixed $k_x=0.1\pi$, $k_y=0$, and $k_-=-\pi/2$, respectively. 
It is seen that the two bands are well separated in both real and imaginary energies only in Fig.~\ref{fig:3}(a), while have crossing points of imaginary energies in Fig.~\ref{fig:3}(b)(c), where the dynamical parameters need to be chosen carefully to avoid jumping of the state between bands~\cite{Silberstein2020}.

For the simplest case with separated imaginary energies in Fig.~\ref{fig:3}(a), we consider an initial wave packet with $\bm{k}_0=(0.1\pi,0)$ and $|u_{\bm{k}_0}\rangle$ the eigenstate of $\mathcal{H}(\bm{k}_0)$ with larger imaginary energy, which is placed at the center of the 2D lattice to avoid the effect of boundary scattering on the dynamics. 
The static forces $(F_x,F_y)=(0,0.25)$ are applied to reproduce the energy spectrum of the energy band, as shown in Fig.~\ref{fig:3}(d).
The rest two cases in Fig. \ref{fig:3}(b) and (c) with fixed $k_x$ and $k_-$ are more sophisticated, as the spectra inevitably contain a pair of crossing points of imaginary eigenenergies at $k_x=\pm\pi/2$.
Nevertheless, with properly chosen initial states and static forces, the full spectrum of the two bands can still be reconstructed through energy trajectories obtained from the wave-packet dynamics, as shown in Fig. \ref{fig:3}(e) and (f).
With more details and examples demonstrated in Supplemental Materials~\cite{SupMat}, we argue that there are two key factors for obtaining the full energy spectrum from this method: 
(i) the difference of loss rate between two bands is small, so states prepared in one band are still able to evolve through a full period of Bloch oscillations before the other band dominates the evolution; and 
(ii) For Fig. \ref{fig:3}(c) and (f), the considered dynamical process has avoided passing the momentum $k_x=\pi/2$ in a Bloch period, where not only the imaginary gap closes, but the real gap takes its minimal value, so that the adiabatic evolution becomes the most vulnerable.

The GDSE in our model is now readily to be read out from our simulation results of the wave-packet dynamics under static forces in different directions.
It is shown in Fig.~\ref{fig:3}(d) and (e) that during one period of Bloch oscillations, the reconstructed complex eigenvalues always go back itself shown by the black arrows, forming an arc on the complex plane, which indicates the absence of NHSE along $y$ and $x$ directions. 
While in Fig.~\ref{fig:3}(f), the reconstructed spectral trajectories go through each eigenvalue once during one period and forms a spectral loop on the complex plane, indicating the skin accumulation along $x+y$ direction.

\emph{{\clr Skin accumulation and dynamical signature toward arbitrary directions}.---}~
In a more generic scenario, a static force can be applied with arbitrary orientations, revealing skin accumulating strength toward different directions of the 2D $sp$ lattice.
Here we consider a lattice rotated clockwise by an angle $\theta$ from a square lattice, as illustrated in Fig.~\ref{figRotate}(a).
Owing to the GDSE, eigenstates are extended in the square lattice with $\theta=0$, 
 but exhibit skin accumulation otherwise, as demonstrated by $\bar{\rho}(\bm{r})$ defined in Eq.(\ref{eq:saden}) for the diamond lattice ($\theta=\pi/4$) Fig.~\ref{fig:1}(d),
 and the one in Fig.~\ref{figRotate}(a) with $\theta=\pi/6$.
Therefore, skin accumulation is expected to emerge and become stronger with $\theta$ continuously increases from $0$ to $\pi/4$.
To see this, 
We specify the boundary section $\mathcal{B}$ as the outermost two layers [i.e. the region between black solid and dashed squares in Fig.~\ref{figRotate}(a)], and define the ratio between the density of modes at boundary and total density of eigenstates as $\bar{\rho}_{\textrm{b}}=\sum_{\bm{r} \in \mathcal{B}} \bar{\rho}(\bm{r})/\sum_{\bm{r}}\bar{\rho}(\bm{r})$. 
$\bar{\rho}_{\textrm{b}}$ under different geometric shapes are shown in Fig.~\ref{figRotate}(b). 
It is clearly shown that the square geometry ($\theta=0$) has a vanishing boundary density ratio, $\rho_{\textrm{b}}\approx0$, which increases with $\theta$ and reaches its maximal at the diamond geometry with $\theta=\pi/4$~\cite{note1}. 

As in the previous case, the NHSE under different open-boundary geometries is expected to be reflected by the spectral trajectories reconstructed from wave-packet dynamics under the applied static force with different orientations, e.g. $\tan\theta=F_x/F_y$ in Fig. \ref{figRotate}(c).
However, it is difficult to reconstruct a complete loop-like spectrum from Bloch oscillations for an arbitrary $\theta$,
since the time evolution may have a much longer period $T$, or even become quasi-periodic for most values of $\theta$.
Alternatively, the emergence of NHSE can also be identified from dynamical degeneracy splitting of the reconstructed energy trajectories~\cite{Kai2022DDS}, even without passing through a full period of Bloch oscillations.
That is, for a given real excitation energy, the disappearing of degeneracy of eigenenergies leads to imbalanced lifetime (i.e., different imaginary parts of energy) for eigenstates with forward and backward velocities,
resulting in a one-way propagation and skin accumulation corresponding to the component with a longer lifetime~\cite{Ashvin2019_PRL}.
As seen in Fig. \ref{figRotate}(c),
the reconstructed energy trajectory shows a stronger splitting when $\theta$ increases from $0$ to $2\pi$,
corresponding to the increased strength of NHSE reflected by Fig. \ref{figRotate}(b).

\emph{{\clr Conclusions and discussion}.---}
To conclude, we have unveiled the emergence of reciprocal GDSE in a 2D double-well optical lattice with atom loss, and proposed to detect it via the anisotropic behavior of non-unitary Bloch oscillations. 
Relying only on bulk dynamics, our scheme provides a convenient way to probe NHSE in cold atoms loaded in optical lattices, where open boundaries for different geometries usually requires more sophisticated techniques to realize, such as programmable digital-micromirror devices~\cite{Ha2015,Gauthie16,Mazurenko2017,Qiu2020}. 
Specifically, skin accumulation in the lattice is identified from spectral winding information or dynamical degeneracy splitting of the energy trajectories reconstructed from the wave-packet dynamics, allowing for probing not only GDSE, but also other types of NHSE as well, e.g. a line NHSE in our model induced by atomic rotation, as discussed in Supplemental Materials~\cite{SupMat}. 
Thus, our work not only opens up an avenue towards experimental realization and detection of the exotic GDSE in a quantum many-body platform, 
but also paves the way toward further exploration on how different types of NHSE interplay with quantum many-body physics.

\newpage

\begin{widetext}

	\setcounter{equation}{0} \setcounter{figure}{0} \setcounter{table}{0} %
	\renewcommand{\theequation}{S\arabic{equation}} \renewcommand{\thefigure}{S%
		\arabic{figure}} \renewcommand{\bibnumfmt}[1]{[S#1]} 
	\renewcommand{\citenumfont}[1]{S#1}

\begin{center}
\textbf{\large Supplementary Materials}
\end{center}

%

\section{Fitting tight-binding parameters from optical potentials}

In the main text, parameters are chosen for a clear demonstration of GDSE, without resorting to the optical lattice potential.
In this section, we give an example of calculation about these parameters with the method reviewed in Ref.~\citep{Li_2016SM}. The static optical potential of our model is given by 
\begin{equation}
V(x,y) = {V_x}{\sin ^2}\frac{2\pi}{\lambda_L}x + {V_1}{\sin ^2}\frac{2\pi}{\lambda_L}y + {V_2}{\sin ^2}(\frac{4\pi}{\lambda_L}y + \phi/2), 
\end{equation}
with $V_x$ ($V_1,V_2$) describing the strength of a static single-well (double-well) potential along $x$ ($y$) direction, and $\phi$ is a phase shift of the double-well potential. 
The minimums of the potential are located at 
$x_0=n\pi /k$, and $y_0\approx n\pi/k$ and $(n+1/2)\pi/k$ for the double wells respectively (with $n$ an arbitrary integer), whose accurate values depend on the potential strengths and are obtained numerically in our calculation.
Expanding the potential to second order at the minimums, we obtain\\
\begin{eqnarray}
V(x,y) &=&   \frac{1}{2}({V_x}{\rm{ + }}{V_1}{\rm{ + }}{V_2}) - \frac{1}{2}[{V_x}\cos (2kx) + {V_1}\cos (2ky) + {V_2}\cos (4ky + \phi )]\nonumber\\
 &&\approx   \frac{1}{2}({V_x}{\rm{ + }}{V_1}{\rm{ + }}{V_2}) - \frac{1}{2}[{V_x}\cos (2k{x_0}) + {V_1}\cos (2k{y_0}) + {V_2}\cos (4k{y_0} + \phi )]\nonumber\\
&&+ \frac{1}{2} \times \frac{1}{2}[4{V_x}{k^2}\cos (2k{x_0})]{(x - {x_0})^2} + \frac{1}{2} \times \frac{1}{2}[4{V_1}{k^2}\cos (2k{y_0}) + 16{V_2}{k^2}\cos (4k{y_0} + \phi )]{(y - {y_0})^2}\nonumber\\
  &&=  \frac{1}{2}({V_x}{\rm{ + }}{V_1}{\rm{ + }}{V_2}) - \frac{1}{2}[{V_x}\cos (2k{x_0}) + {V_1}\cos (2k{y_0}) + {V_2}\cos (4k{y_0} + \phi )]\nonumber\\
 &&+ \frac{m}{2}[4{V_1}\frac{{{k^2}}}{{2m}}\cos (2k{y_0}) + 16{V_2}\frac{{{k^2}}}{{2m}}\cos (4k{y_0} + \phi )]{(y - {y_0})^2}\nonumber\\
 &&+ \frac{m}{2}[4{V_x}\frac{{{k^2}}}{{2m}}\cos (2k{x_0})]{(x - {x_0})^2},
\end{eqnarray}
with $k=\frac{2\pi}{\lambda_L}$. Thus, the total Hamiltonian can be written as
\begin{equation}
H = \frac{{p_x^2 + p_y^2}}{{2m}} + V(x,y) = \frac{{p_x^2 + p_y^2}}{{2m}} + \frac{m}{2}\omega _x^2\delta _x^2 + \frac{m}{2}\omega _y^2\delta _y^2,
\end{equation}
with ${\delta _x} = x - {x_{0 }}$, ${\delta _y} = y - {y_{0 }}$, and
\begin{equation}
{\omega _x} = \sqrt {4{V_x}\frac{{{k^2}}}{{2m}}\cos (2k{x_0})}  = \sqrt {4{V_x}{E_r}\cos (2k{x_0})} ,~~
{\omega _y} = \sqrt {4{V_1}{E_r}\cos (2k{y_0}) + 16{V_2}{E_r}\cos (4k{y_0} + \phi )} ,
\end{equation}
where $E_r=\hbar^2k^2/2m$. The energy and wavefunctions are given by 
\[{E_{{l_{x,}}{l_y}}} = \hbar \left[ {{\omega _x}({l_x} + \frac{1}{2}) + {\omega _y}({l_y} + \frac{1}{2})} \right] + V({x_{0 }},{y_{0 }}),\]
\[{\psi _{{l_x},{l_y}}} = \sqrt {\frac{1}{{{2^{{l_x} + {l_y}}}{l_x}!{l_y}!}}} {\left( {\frac{1}{{2\pi }}} \right)^{1/2}}{\left( {{\omega _x}{\omega _y}} \right)^{1/4}}\exp \left[ { - \frac{1}{4}\left( {{\omega _x}\delta _x^2 + {\omega _y}\delta _y^2} \right)} \right]{H_{{l_x}}}\left( {\sqrt {\frac{{{\omega _x}}}{2}} \delta _x^{}} \right){H_{{l_y}}}\left( {\sqrt {\frac{{{\omega _y}}}{2}} \delta _y^{}} \right),\]
where ${l_{x/y}}$ represent quantum numbers along $x$ and $y$ directions respectively, and 
\[
{H_l}(z) = {( - 1)^l}{e^{{z^2}}}\frac{{{d^l}}}{{d{z^l}}}\left( {{e^{ - {z^2}}}} \right)
\]
 is the Hermite polynomials.  This wavefunction gives the zero order approximation of the Wannier state of the potential. Here the $s$ orbital is given by ($l_x$, $l_y$) = (0, 0),
and the $p_x$ orbital is ($l_x$, $l_y$) = (1, 0). 
It should be noted that the $s$ and $p_x$ orbitals in our consideration correspond to the double wells with different depths and different locations. Thus this zero order approximation is not accurate enough due to the non-zero overlap integration for different bases of them. Thus we adapt the method from  Ref.~\citep{Li_2016SM} to construct Wannier states with a small overlap integration. 
Explicitly, the zero order harmonic oscillator wavefunctions are given by $\phi_{\alpha}(\mathbf{x}-\mathbf{R})$ lozalized on sites $\mathbf{R}$, which are nearly orthogonal, i.e. 
\begin{equation}
\int \mathrm{d} \mathbf{x} \phi_\alpha^*(\mathbf{x}-\mathbf{R}) \phi_{\alpha^{\prime}}\left(\mathbf{x}-\mathbf{R}^{\prime}\right)=\delta_{\alpha \alpha^{\prime}} \delta_{\mathbf{R} \mathbf{R}^{\prime}}+\epsilon_{\alpha \mathbf{R}, \alpha^{\prime} \mathbf{R}^{\prime}},
\end{equation}
with $\epsilon_{\alpha \mathbf{R}, \alpha^{\prime} \mathbf{R}^{\prime}}$ being approximately zero. 
Therefore we can improve the orthogonality by considering a new basis
\begin{equation}
\tilde{\phi}_\alpha(\mathbf{x}-\mathbf{R})=\phi_\alpha(\mathbf{x}-\mathbf{R})-\frac{1}{2} \sum_{{ \alpha}^{\prime} \mathbf{R}^{\prime}} \epsilon_{\alpha^{\prime} \mathbf{R}^{\prime}, \alpha \mathbf{R}} \phi_{\alpha^{\prime}}\left(\mathbf{x}-\mathbf{R}^{\prime}\right),
\end{equation}
which is normalized through
\begin{equation}
\tilde{\phi}_\alpha(\mathbf{x}) \rightarrow \tilde{\phi}_\alpha(\mathbf{x}) / \sqrt{\int \mathrm{d} \mathbf{x}^{\prime}\left|\tilde{\phi}_\alpha\left(\mathbf{x}^{\prime}\right)\right|^2}.
\end{equation}
Thus this new set of wavefunctions shall have smaller overlaps between each other,
\begin{equation}
\int \mathrm{d} \mathbf{x} \tilde{\phi}_\alpha^*(\mathbf{x}-\mathbf{R}) \tilde{\phi}_{\alpha^{\prime}}\left(\mathbf{x}-\mathbf{R}^{\prime}\right)=\delta_{\alpha \alpha^{\prime}} \delta_{\mathbf{R} \mathbf{R}^{\prime}}+\mathcal{O}\left(\epsilon^2\right).
\end{equation}
Performing the iteration procedure $N$ times, an orthogonal normal basis with  $\mathcal{O}\left(\epsilon^{2^N}\right)$ precision can be reached. After obtaining the orthogonal basis, tunnelings parameter between $\mathbf{R}$ and $\mathbf{R}^{\prime}$ are calculated as
\begin{equation}
t_{\alpha \alpha^{\prime}}\left(\mathbf{R}-\mathbf{R}^{\prime}\right)=\int \mathrm{d} \mathbf{x} \tilde{\phi}_\alpha^*(\mathbf{x}-\mathbf{R}) H(\mathbf{x}) \tilde{\phi}_{\alpha^{\prime}}\left(\mathbf{x}-\mathbf{R}^{\prime}\right),
\end{equation}
where $H(\mathbf{x})$ has the form of
\begin{equation}
H(\mathbf{x})=-\frac{\hbar^2}{2 m} \vec{\nabla}^2+V(x,y) .
\end{equation}

As an example, we set $V_1=10$ and $V_2=13, $ for different values of $ \phi$ and $V_x$, and the corresponding tight-binding parameters are given in Table.~\ref{TabT},
which are comparable to our selected parameters in the main text. 
To see the GDSE, we plot  in ~Fig.\ref{figSM1}(a) and (b) the density distribution and spectral winding along $k_+=k_x+k_y$ direction for one set of the parameters from Table.~\ref{TabT}.
The skin boundary localization is weaker than the example in the main text, yet the emergence of skin accumulation can still be predicted from the nontrivial spectral winding in Fig.\ref{figSM1}(b).

\begin{figure*}
\centering
\includegraphics[width=10.0cm]{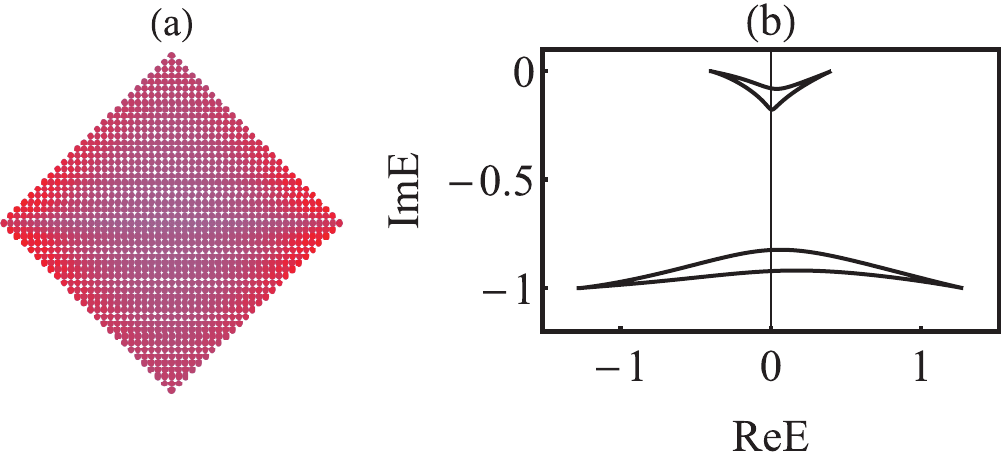}
\caption{\label{figSM1}
 \textbf{Average density distribution and spectral winding along $k_+=k_x+k_y$ direction.} (\textbf{a}) Average density distribution on a diamond geometry, which shows weak localization on the boundaries. (\textbf{b}) Complex energy spectrum with nontrivial spectral winding along $k_+=k_x+k_y$ direction, with fixed $k_-=k_x-k_y=\pi/2$. 
 Parameters are $t_s=-0.2, t_p=0.64, t_{sp}=0.16, t_{sp}^{\prime}=0.04$. The diagonal length of the diamond shape is $L=61$. 
}
\end{figure*}

\begin{table}
\begin{center}
\begin{tabular}{|c|c|cccc|} \hline
$\phi$ & $ V_x$ & $t_s$ &  $t_p$ &  $t_{sp}$ & $t_{sp}^{\prime}$\\ \hline
$0.8\pi$ & 0.8 & -0.2 & 0.64 & 0.16 &0.04\\ \hline
$0.8\pi$ & 1.2 & -0.18 & 0.62 & 0.13 & 0.03\\ \hline
$0.85\pi$ &1&-0.19 & 0.63 & 0.14 & 0.03\\ \hline
\end{tabular}
\caption{Hopping parameters versus $\phi$ for the tight-binding model, in unit of $E_r=\hbar^2 k^2/2m$.}
\label{TabT} 
\end{center}
\end{table}


\section{The volume law of geometry-dependent skin effect}

The volume law of GDSE states that the number of boundary (skin) modes is proportional to the volume of the lattice, $N_{b}\propto V$. To confirm this statement, we define boundary modes as those with their summed density at the selected boundary $\rho_b>0.5$,
with
$$\rho_b:=\sum_{\bm{r} \in \mathcal{B}}\sum_{\alpha=s,p_x}|\psi_{n,\alpha,\bm{r}}|^2$$
and $\mathcal{B}$ the selected boundary (defined as the outermost two layers of lattice sites).
Note that this definition does not distinguishes non-Hermitian skin modes from conventional edge states raised from Hermitian topology.
The results for verifying the volume law of GDSE are shown in Fig.\ref{figs7} (a) and (b). 
\begin{figure*}[t]
	\begin{center}
		\includegraphics[width=.6\linewidth]{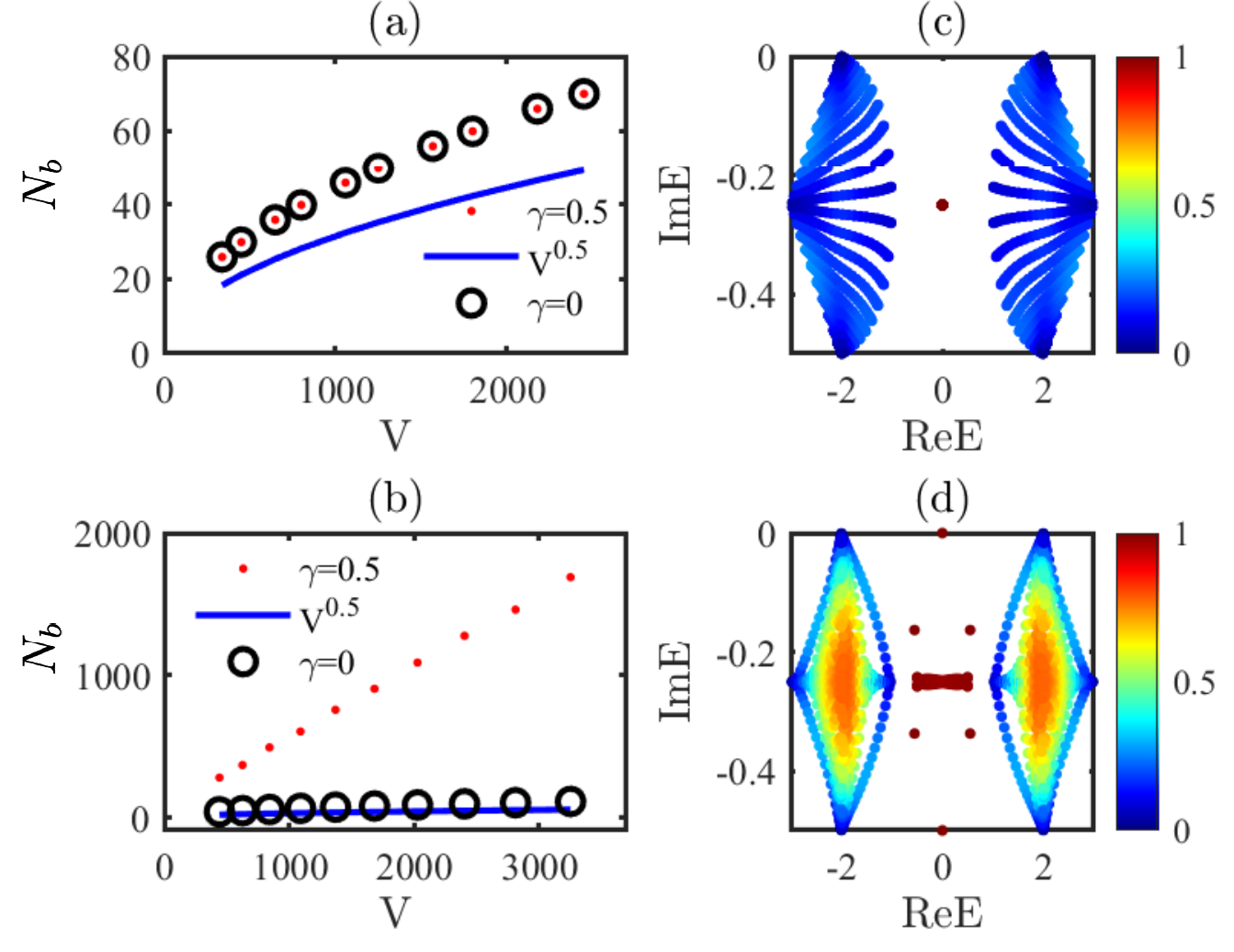}
		\par\end{center}
	\protect\caption{\label{figs7} 
		\textbf{ The area law and localization properties of non-Hermitian $sp$ ladder model.}
		(\textbf{a}) and (\textbf{b}) Number of boundary modes $N_{b}$ versus the system's volume in square and diamond geometries respectively. Red points (black circles) correspond to boundary modes with $\gamma=0.5$ ($\gamma=0$). Blue line is given by $N_b=\sqrt{V}$ as a comparison. 
		Boundary modes are defined as eigenstates with their density at outermost two layers of lattice sites $\rho_b>1/2$.
		(\textbf{c}) and (\textbf{d}) OBC spectra with eigenstates marked by different colors according to $\rho_b$, square geometry and diamond geometries respectively. The size of square geometry and diamond geometry are $L_{x}\times L_{y}=25\times 25$ ($V=625$) and  $L_{x}\times L_{y}=39\times 39$ ($V=761$) respectively. $L_{x,y}$ represents the diagonal length for diamond geometry. Other parameters are $ t_{s}=1,t_{p}=1,t_{sp}=1,t_{sp}^{\prime}=0.5$.
	} 
\end{figure*}
In square geometry, $N_{b}$ scales similarly as $\sqrt{V}$, and shows no differences between Hermitian and non-Hermitian cases (with zero and nonzero $\gamma$ respectively).
Thus we can conclude that the edge states here are originated from Hermitian topological properties.
In contrast, for diamond geometry, $N_{b}$ becomes much larger and is proportional to the volume of system when non-Hermiticity is turned on,
indicating the appearance of non-Hermitian skin modes.
Therefore, we verify the volume law of 2D $sp$ ladder model numerically. 
In Fig.\ref{figs7} (c) and (d), we demonstrate the complex spectra of the system under different geometries, with different colors revealing the boundary density $\rho_b$ of each eigenstate.
It is clearly seen that only some in-gap zero-energy eigenstates are localized at the system's boundary (with $\rho_b\approx1$) in square geometry,
whereas most eigenstates become boundary modes (with smaller $\rho_b$, but still larger than $0.5$) in diamond geometry.

\section{Non-Hermitian skin effect along oblique directions in the $sp$ ladder model}
The Bloch Hamiltonian is given by 
\begin{equation}
  \mathcal{H}({k_x},{k_y}) = \boldsymbol{h}(k_x,k_y) \cdot \boldsymbol{\sigma}-i\gamma(\sigma_0-\sigma_z)/2 ,
\end{equation}
where $\bm{\sigma}$ is a vector of the $2\times2$ identity matrix and Pauli matrices, and $\bm{h}(k_x,k_y)$ is a vector with four components, ${h_0}(k_x,k_y) = ({t_p} - {t_s})\cos {k_x}$, 
${h_x}({k_x},{k_y}) =  - 2t'_{sp}\sin {k_y}\sin {k_x}$, 
$h_y(k_x,k_y) = 2 t_{sp} \sin{k_x} + 2 t^{\prime}_{sp} \cos{k_y} \sin{k_x}$, ${h_z}(k_x,k_y) = -(t_p + t_s)\cos{k_x}$. 
To see the NHSE along oblique directions,
we define
\[\left\{ \begin{array}{l}
{k_ + } = \frac{{{k_x} + {k_y}}}{2},\\
{k_ - } = \frac{{{k_x} - {k_y}}}{2}.
\end{array} \right. \Leftrightarrow \left\{ \begin{array}{l}
{k_x} = {k_ + } + {k_ - },\\
{k_y} = {k_ + } - {k_ - },
\end{array} \right.\]
with $k_+$ and $k_-$ the crystal momenta along $x+y$ and $x-y$ directions.
Thus the Hamiltonian can be written as 
\begin{equation}
  \mathcal{H}({k_{+}},{k_{-}}) = \boldsymbol{h}(k_{+},k_{-}) \cdot \boldsymbol{\sigma}-i\gamma(\sigma_0-\sigma_z)/2 ,
\end{equation}
with
\[\begin{array}{l}
{h_0}({k_ + },{k_ - })= ({t_p} - {t_s})\cos ({k_ + } + {k_ - }) = ({t_p} - {t_s})\cos {k_ + }\cos {k_ - } - ({t_p} - {t_s})\sin {k_ + }\sin {k_ - },\\
{h_x}({k_ + },{k_ - }) =  - 2t'_{sp}\sin ({k_ + } - {k_ - })\sin ({k_ + } + {k_ - }) = t'_{sp}\cos (2{k_ + }) - t'_{sp}\cos (2{k_ - }),\\
 {h_y}({k_ + },{k_ - }) =  2{t_{sp}}\sin ({k_ + } + {k_ - }) + 2{{t'}_{sp}}\cos ({k_ + } - {k_ - })\sin ({k_ + } + {k_ - })\\
 =  2{t_{sp}}\sin {k_ + }\cos {k_ - } + 2{t_{sp}}\cos {k_ + }\sin {k_ - } + {{t'}_{sp}}\sin (2{k_ + }) + {{t'}_{sp}}\sin (2{k_ - }),\\
{h_z}({k_ + },{k_ - }) =  - ({t_p} + {t_s})\cos ({k_ + } - {k_ - }) =  - ({t_p} + {t_s})\cos {k_ + }\cos {k_ - } - ({t_p} + {t_s})\sin {k_ + }\sin {k_ - }.
\end{array}\]

Rewriting the Hamiltonian with $\beta_+=\exp(ik_+)$ and  $\beta_-=\exp(ik_-)$, one obtains
\[H(\beta_+,\beta_-) = \left( {\begin{array}{*{20}{c}}
{ - \beta_+\beta_-{t_s} - \frac{{{t_s}}}{{\beta_+\beta_-}}}&{ - {{t'}_{sp}}{\beta_-^2} - {t_{sp}}\beta_+\beta_-  + \frac{{{{t'}_{sp}}}}{{{\beta_+^2}}} + \frac{{{t_{sp}}}}{{\beta_+\beta_-}}}\\
{{{t'}_{sp}}{\beta_+^2} + {t_{sp}}\beta_+\beta_-   - \frac{{{t_{sp}}}}{{\beta_+\beta_-}} - \frac{{{{t'}_{sp}}}}{{{\beta_-^2}}}}&{\beta_+\beta_-{t_p} + \frac{{{t_p}}}{{\beta_+\beta_-}} - i\gamma }
\end{array}} \right)\]
We then view $\beta_-$ as a parameter and write down a polynomial $f(\beta_+,E)=\det[H(\beta_+,\beta_-)-E]$ as 
\[\begin{array}{l}
f(\beta_+,E(\beta_-))={\beta_+^3}{t_{sp}}{{t'}_{sp}}\beta_- + {\beta_+^2}\left( { - {t_p}{t_s}{\beta_-^2} + t_{sp}^2{\beta_-^2} + {{t'}_{sp}}^2{\beta_-^2} } \right)\\
\beta_+\left( { - E{t_p}\beta_- + E{t_s}\beta_- + i\gamma {t_s}\beta_- + {t_{sp}}{{t'}_{sp}}^2{\beta_-^3} - \frac{{2{t_{sp}}{{t'}_{sp}}}}{\beta_-} } \right)\\
  + {E^2} + i\gamma E - 2{t_p}{t_s} - 2t_{sp}^2 - 2{{t'}_{sp}}^2 \\
 + \frac{1}{\beta_+}\left( { - \frac{{E{t_p}}}{\beta_-} + \frac{{E{t_s}}}{\beta_-} + \frac{{i\gamma {t_s}}}{\beta_-} + \frac{{{t_{sp}}{{t'}_{sp}}}}{{{\beta_-^3}}} - 2{t_{sp}}{{t'}_{sp}}\beta_-} \right)\\
 + \frac{1}{{{\beta_+^2}}}\left( { - \frac{{{t_p}{t_s}}}{{{\beta_-^2}}} + \frac{{t_{sp}^2}}{{{\beta_-^2}}} + \frac{{{{t'}_{sp}}^2}}{{{\beta_-^2}}}} \right)\\
 + \frac{{{t_{sp}}{{t'}_{sp}}}}{{{\beta_+^3}\beta_-}}.
\end{array}\]
Sorting the six roots of $f(\beta_+,E(\beta_-))=0$ by their absolute values, 
the GBZ solution of this model, which describes skin accumulation along $x+y$ direction, is given by the third and fourth roots when they have the same absolute values.
Similarly, by taking $\beta_+$ as a parameter, we can rewrite the polynomial as
%
%
\[\begin{array}{l}
g(\beta_-,E(\beta_+))={\beta_-^3}{t_{sp}}{{t'}_{sp}}\beta_+ + {\beta_-^2}\left( { - {t_p}{t_s}{\beta_+^2} + {t_{sp}}^2{\beta_+^2} + {{t'}_{sp}}^2{\beta_+^2} } \right)\\
 + \beta_-\left( { - E{t_p}\beta_+ + E{t_s}\beta_+ + i\gamma {t_s}\beta_+ + {t_{sp}}{{t'}_{sp}}{\beta_+^3} - \frac{{2{t_{sp}}{{t'}_{sp}}}}{\beta_+} } \right)\\
 - 2{t_p}{t_s} - 2{t_{sp}}^2 - 2{{t'}_{sp}}^2   + {E^2} + i\gamma E\\
 + \frac{1}{\beta_-}\left( { - \frac{{E{t_p}}}{\beta_+} + \frac{{E{t_s}}}{\beta_+} + \frac{{i\gamma {t_s}}}{\beta_+} + \frac{{{t_{sp}}{{t'}_{sp}}}}{{{\beta_+^3}}} - 2{t_{sp}}{{t'}_{sp}}\beta_+ } \right)\\
 + \frac{1}{{{\beta_-^2}}}\left( { - \frac{{{t_p}{t_s}}}{{{\beta_+^2}}} + \frac{{{t_{sp}}^2}}{{{\beta_+^2}}} + \frac{{{{t'}_{sp}}^2}}{{{\beta_+^2}}} } \right) + \frac{{{t_{sp}}{{t'}_{sp}}}}{{\beta_+{\beta_-^3}}},
\end{array}\]
with $f(A,E(B))=g(A,E(B))$.
Thus, 
the GBZ solutions (and skin accumulation) along $x+y$ and $x-y$ are equivalent, which is related to reciprocity-mirror symmetry as indicated in the main text.

\section{Non-unitary dynamics in Bloch oscillations and non-Hermitian skin effect}
Skin effect is related to the spectral topology by a winding number. To measure this spectral winding number of complex eigenenergies, we consider a method using Bloch oscillations for reconstructing the energy spectrum. This method was first adopted to deal with a two-dimensional (2D) Hatano-Nelson model~\cite{Gong2018sSM}.  
Extending this method to two dimensions, we can study higher-dimensional skin effects. 
Explicitly, with a static force $\boldsymbol{F}$ applied to the system,
a wave-packet shall oscillate in the lattice periodically (known as the Bloch oscillations),
which can be described by the classical equation of motions~\cite{Longhi2009SM,Longhi2015_PRASM,Graefe2016SM,Gong2018_PRXSM}
\begin{equation}
\frac{d\boldsymbol{k}}{dt} =\boldsymbol{F},{\rm{ }}\frac{d\langle\boldsymbol{r}\rangle_t}{dt} = {\mathop{\rm Re}\nolimits} {\nabla _{\boldsymbol{k}}}E(\boldsymbol{k}),\frac{{d{\rm{ln(}}{{\cal N}_t}{\rm{)}}}}{{dt}} = 2{\mathop{\rm Im}\nolimits} E(\boldsymbol{k}),\label{Ss2}
\end{equation}
where  ${{\cal N}_t}$ and $\langle\boldsymbol{r}\rangle_t$ are the norm and center of mass of the instantaneous state at time $t$ respectively. 
Thus we can reconstruct the complex spectrum along a specific path in the Brillouin zone from the motion of the wave-packet.

\begin{figure*}
\centering
\includegraphics[width=10.0cm]{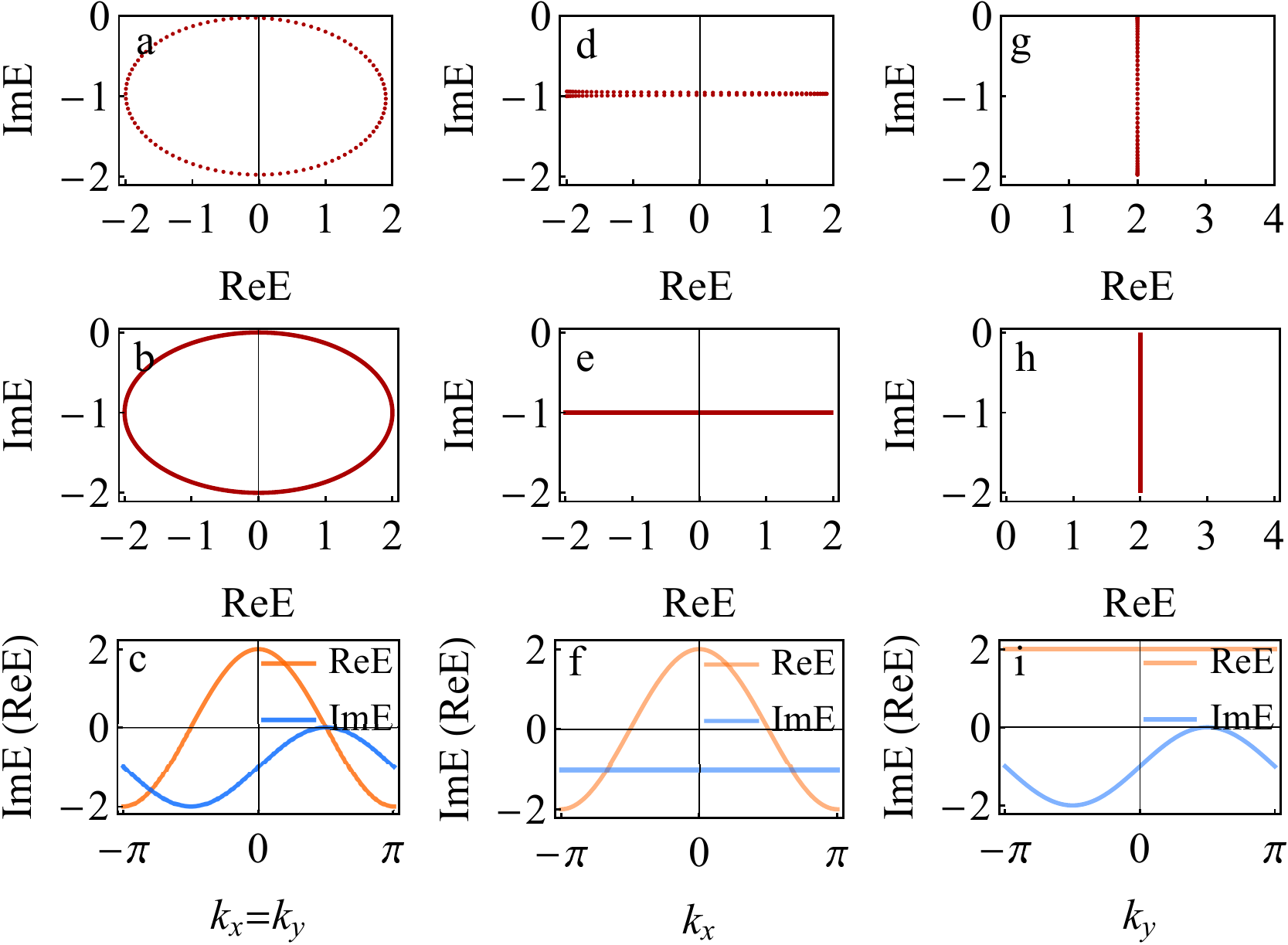}
\caption{\label{figS1}\textbf{Comparison between dynamically reconstructed energy trajectories and complex energy spectra of the 2D Hatano-Nelson model.} \textbf{a}) Reconstructed energy trajectory (a loop) with ${{F}_{x}}={{F}_{y}}=10$, where the wave-packet has it momentum varying along ${{k}_{x}}={{k}_{y}}$ direction. (\textbf{b-c}) The energy spectrum along ${{k}_{x}}={{k}_{y}}$ in the Brillouin zone. 
(\textbf{d}) Reconstructed energy trajectory (an arc) with ${{F}_{x}}=10,{{F}_{y}}=0$, with the momentum the 
wave-packet varying along $k_x$ direction. (\textbf{e-f}) The energy spectrum along ${{k}_{y}}=0$. (\textbf{g}) Reconstructed energy trajectory (an arc) ${{F}_{x}}=0,{{F}_{y}}=10$, with the momentum the wave-packet varying along $k_y$ direction. (\textbf{h-i}) The energy spectrum along ${{k}_{x}}=0$. The system's size is chosen to be $L_x=L_y=33$. }
\end{figure*}
\subsection{A single-band case in two dimensions}
As a simple example, we first consider a 2D single band model which shows geometry dependent skin effect. The model Hamiltonian is given by \cite{Zhang2022sSM}
\begin{equation}
 H(\boldsymbol{k}) = 2\cos {k_x} + i\sin {k_y} - i. \label{S1}
\end{equation}
To dynamically reconstruct its spectrum, we consider a gaussian wave packet as the initial state,
$$\left| {{\psi _0}} \right\rangle  = A\exp \left[ { - {{\left( {x - {x_0}} \right)}^2}/W - {{\left( {y - {y_0}} \right)}^2}/W}+ik_x^0x+ik_y^0y \right]$$  
with ${{k}_{x}^0}={{k}_{y}^0}=0$, where $A$ is the normalization factor,
and the parameter $W$ controls the width of the wave-packet.
The results obtained from Eq.~\eqref{Ss2} with different statice force are shown in Fig.\ref{figS1} (a), (d) and (g), which accurately reproduce the loop-like spectrum along ${x}+{y}$ direction [Fig.\ref{figS1} (b)], and arc-like spectrum along each of ${x}$ and ${y}$ directions [Fig.\ref{figS1} (e) and (h)]. 
The real and imaginary parts of these spectra are displayed separately in Fig.\ref{figS1} (c), (f) and (i) for a better illustration of the band structures.

In order to further study the effect of finite width of wave-packet , we calculate the trajectory along $x$ direction with different wave packet widths or forces. 
As shown in Fig.\ref{figS6} (a), the reconstructed energy trajectory is converging toward the theoretical one as the widths of the wave-packet $W$ increases.
That is because that a larger wave-packet width corresponds to narrower spreading in momentum space.  
On the other hand, the period of Bloch oscillation is inversely proportional to the strength of the applied force, thus a stronger force corresponds to the shorter oscillation time and prevents the spreading of the wave packet. Consistently,  the reconstructed energy trajectory is converging toward the theoretical one when increasing the force $F_x$, as shown in Fig.\ref{figS6} (b).
\begin{figure*}
\centering
\includegraphics[width=10.0cm]{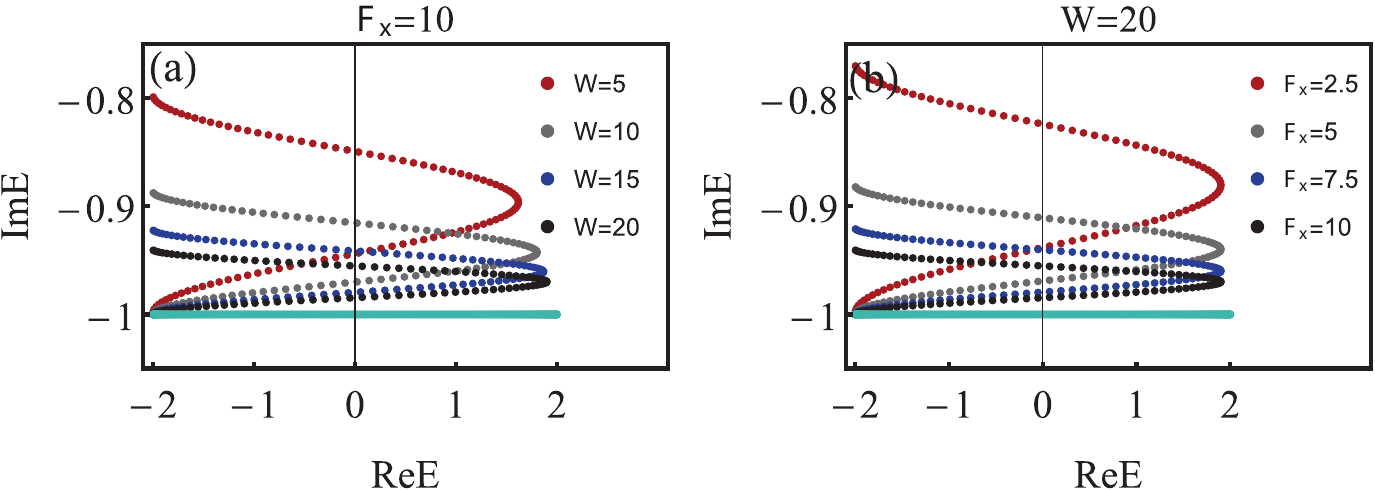}
\caption{\label{figS6}
 \textbf{Reconstructed energy trajectories with $F_y=0$ with different widths of the initial wave-packet and different static forces.}. 
 (\textbf{a}) 
 The reconstructed energy trajectories is  converging toward the energy spectrum when increasing $W$.
 (\textbf{b}) 
 The reconstructed energy trajectories is converging toward the energy spectrum (cyan line) when increasing $F_x$.
$F_y=0$ is set for both panels, and the dynamical results (dots) are expected to reconstruct the spectrum along $k_x$ with $k_y=0$ (cyan line).
}
\end{figure*}

\subsection{A two-band example in one dimension}
Generalizing this method to multiband models is a non-trivial task due to the breaking down of adiabatic theorem~\cite{Silberstein2020SM}. As an example, here we consider a generalized one-dimensional Su-Schrieffer-Heeger (SSH) Hamiltonian \cite{Kunst2018sSM,Zhu2014sSM,Lee2016sSM,Lieu2018SM},
  \begin{equation}
H(k)=\left( \begin{matrix}
   0 & {{t}_{L}}+{{{{t}'}}_{R}}{{e}^{-ik}}  \\
   {{t}_{R}}+{{{{t}'}}_{L}}{{e}^{ik}} & 0  \\
\end{matrix} \right).
  \end{equation}
Note that we have neglected the modifications due to berry connection~\cite{Silberstein2020SM}, which does not affect the spectral winding in our numerical calculations. To dynamically reconstruct its spectrum, we choose the initial state as $\psi (t=0)=u_{{{k}_{0}}}^{+}\exp (i{{k}_{0}}r)\exp \left( -\frac{{{(r-{{r}_{c}})}^{2}}}{2{{\sigma }^{2}}} \right)$, where 
${{r}_{c}}=L/2$ , $\sigma =10$ and $u_{{{k}_{0}}}^{+}$ the eigenstate of $H({{k}_{0}})$ with larger imaginary eigenvalue. The choices of initial momentum $k_0$ in following discussing of specific examples.

As shown in Fig. \ref{figS2}(a), this model has two separable bands forming two arcs in the complex energy plane in the parameter regime of $t_Lt_L'=t_Rt_R'$, where its eigenenergies satisfy $E(k)=E(\pi-k)$. On the other hand, as shown in Fig.  \ref{figS2}(b), their imaginary energies cross at $k_x=0$ and $\pi$, and the dynamics in different momentum intervals between the crossing points is expected to be dominated by the energy band with larger imaginary energy. In Fig. {figS2}(c) and (d), we display the energy spectrum reconstructed from wave-packet dynamics with $k_0$ chosen to be $0$ and $-\pi$ for two different dynamical processes. The results indeed recover the part of the complex spectrum with larger imaginary energies. 
The reconstructed trajectories  passes through the same energy point twice, indicating a zero spectral winding number for the full spectrum and the absence of imaginary energy splitting, both suggesting the absence of NHSE.

In Fig. \ref{figS3}, we consider another parameter regime of the generalized SSH model, with the presence of NHSE. Its spectrum forms a single loop in the complex energy plane [Fig. \ref{figS3}(a)], and the imaginary energies of the two non-separable bands crosses only once in the Brillouin zone [Fig. \ref{figS3}(b)].
Therefore we shall be able to recover half of the single-loop spectrum, as shown in [Fig. \ref{figS3}(c) and (d)]. We can infer the presence of NHSE from the fact that the reconstructed energy trajectory only passes through each point in complex energy plane once during the time evolution.

Finally, we illustrate another example with separable bands and NHSE in Fig. \ref{figS4}.
In this case, one of the two energy bands always have larger imaginary energies and possesses a loop-like spectrum with spectral nontrivial winding.
The reconstructed energy trajectory is capable to capture this signature, unveiling the presence of NHSE in this case.

\begin{figure*}
\centering
\includegraphics[width=11.0cm]{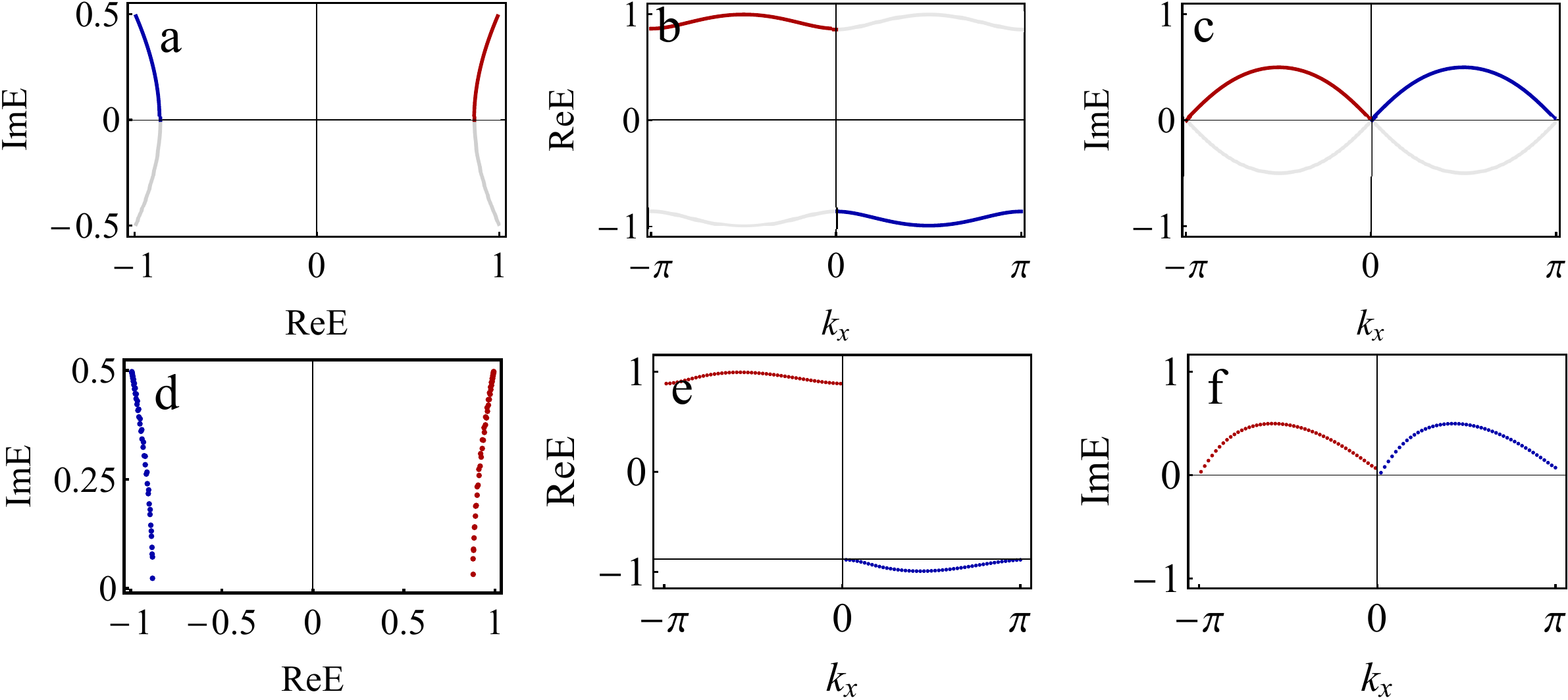}
\caption{\label{figS2}
\textbf{Energy spectrum and its dynamical reconstruction for non-Hermitian SSH model without NHSE.} 
(\textbf{a}) Separable arc-like spectrum on the complex energy plane.
Red (blue) arc corresponds to eigenenergies with larger imaginary parts for $k\in (-\pi ,0]$ ($k\in (0,\pi]$).
(\textbf{b}) real and (\textbf{c}) imaginary eigenenergies as functions of $k_x$.
(\textbf{d}) to (\textbf{f}) dynamical reconstruction of the eigenenergies in (\textbf{a}) to (\textbf{c}) respectively.
Parameters are $L=400$, ${{t}_{L}}=12/11,{{t}_{R}}=10/11$,  ${{{t}'}_{L}}=-5/11$,${{{t}'}_{R}}=6/11$, and the static force is $F=0.1$. 
The initial momentum is ${{k}_{0}}=0$ for the blue dots and $k_0=-\pi$ for the red dots.
}
\end{figure*}
\begin{figure*}
\centering
\includegraphics[width=8.0cm]{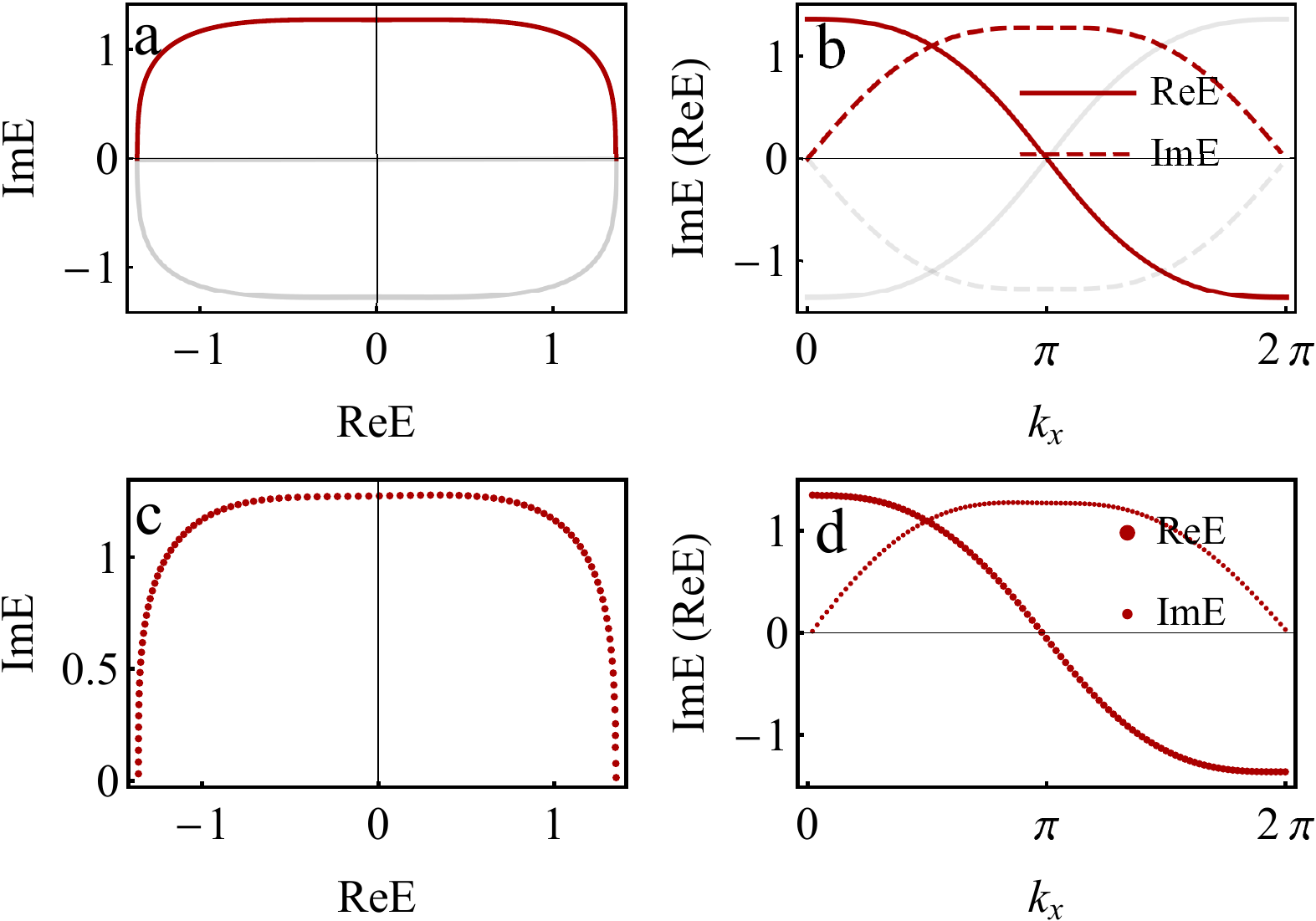}
\caption{\label{figS3}
\textbf{Energy spectrum and its dynamical reconstruction for non-Hermitian SSH model with NHSE, case I.} 
(\textbf{a}) Non-separable lopp-like spectrum on the complex energy plane.
Red arc corresponds to eigenenergies with larger imaginary parts for $k\in (0,2\pi]$.
(\textbf{b}) real and imaginary eigenenergies as functions of $k_x$.
(\textbf{c}) and (\textbf{d}) dynamical reconstruction of the eigenenergies in (\textbf{a}) and (\textbf{b}) respectively.
Parameters are $L=400$, ${{t}_{L}}=1.3$, ${{t}_{R}}=0.7$, ${{{t}'}_{L}}=1.6$, ${{{t}'}_{R}}=-0.5$, and the static force is $F=0.1$. 
The initial momentum is ${{k}_{0}}=0$. 
}
\end{figure*}

\begin{figure*}
\centering
\includegraphics[width=8.0cm]{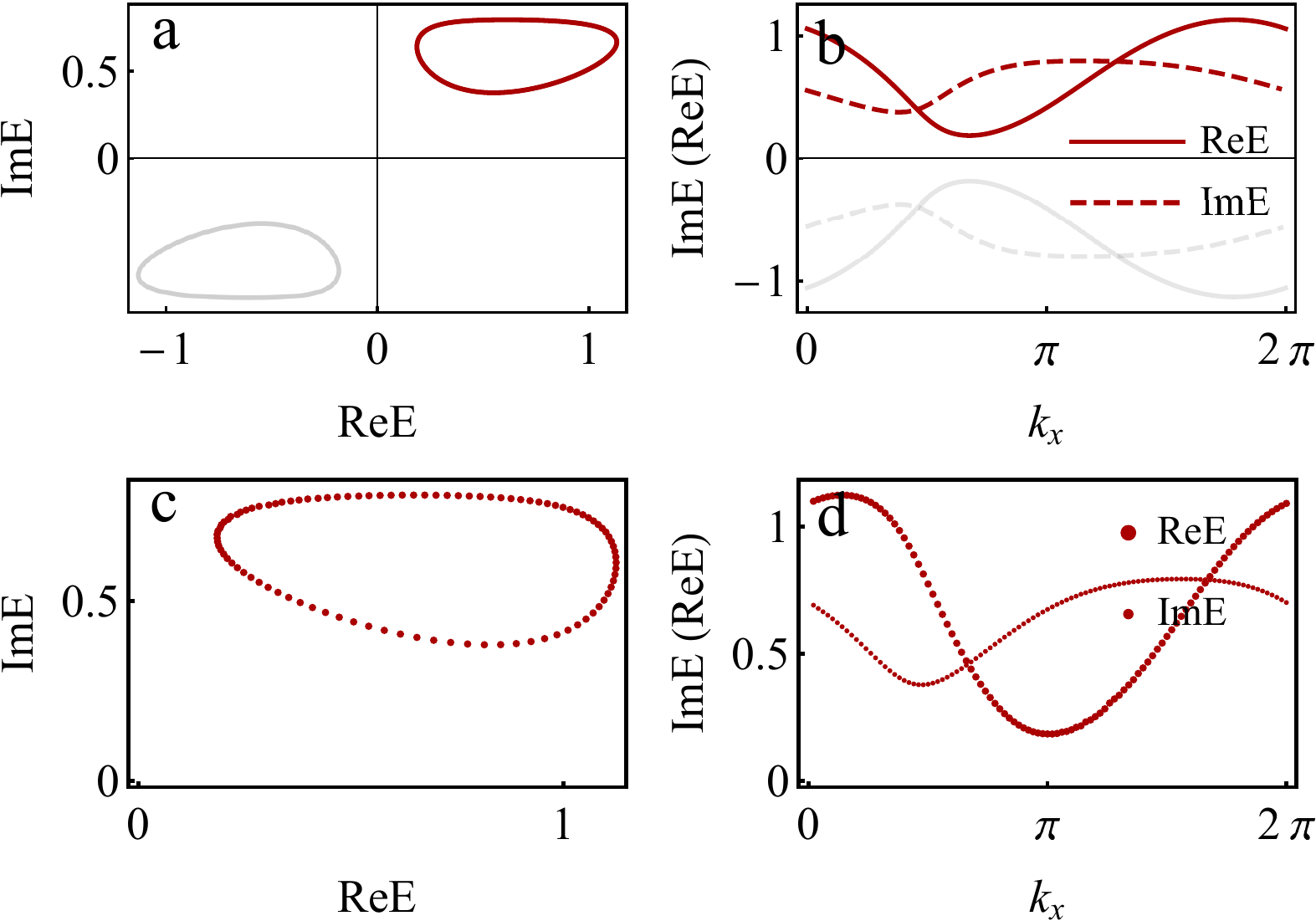}
\caption{\label{figS4}
\textbf{Energy spectrum and its dynamical reconstruction for non-Hermitian SSH model with NHSE, case II.} 
(\textbf{a}) Separable loop-like spectrum on the complex energy plane.
Red arc corresponds to eigenenergies with larger imaginary parts for $k\in (0,2\pi]$.
(\textbf{b}) real and imaginary eigenenergies as functions of $k_x$.
(\textbf{c}) and (\textbf{d}) dynamical reconstruction of the eigenenergies in (\textbf{a}) and (\textbf{b}) respectively.
Parameters are $L=400$, ${{t}_{L}}=1.3i$, ${{t}_{R}}=0.7$, ${{{t}'}_{L}}=0.2$, ${{{t}'}_{R}}=0.9$ and the static force is $F=0.025$. 
The initial momentum is ${{k}_{0}}=3\pi/4$. 
}
\end{figure*}

\section{The dynamics in multiple periods of Bloch oscillations}

In the main text, we have used the same definition of Bloch oscillation period as for Hermitian cases (i.e. $T_B=2\pi/F$ for one dimension). However, there is a subtle difference for non-Hermitian cases due to the decay effect and the breakdown of adiabatic theorem. 
In this section we will explain the differences between Bloch oscillations in Hermitian and non-Hermitian systems, and demonstrate more details of the latter case for our 2D $sp$ ladder model in the main text.

In Hermitian systems, Bloch oscillations occur when a static force is applied to the system, 
leading to a periodic evolution of a wave-packet, 
which either always stay in one energy band (i.e. an adiabatic process when the force is relatively weaker than the band gap), 
or jump between different bands (i.e. a non-adiabatic process with a stronger force).
In non-Hermitian cases, possible particle loss leads to two consequences that will affect the time evolution: (i) the decay effect, i.e., the wave-packet shall decay and eventually vanish during the time evolution; and (ii) the breakdown of adiabatic theorem, i.e., energy band with the largest imaginary energies corresponds to the weakest decay rate, resulting in another type of jumping of the wave-packet between different bands.
Noticing that both these two effects require a certain amount of time to manifest,
we need to properly chose parameters of the applied static force and non-Hermitian loss, so that the time evolution is slow enough to fulfill the adiabatic condition,
but not too slow comparing to the loss rate of the system.
 \begin{figure*}[t]
 	\begin{center}
\includegraphics[width=0.6\linewidth]{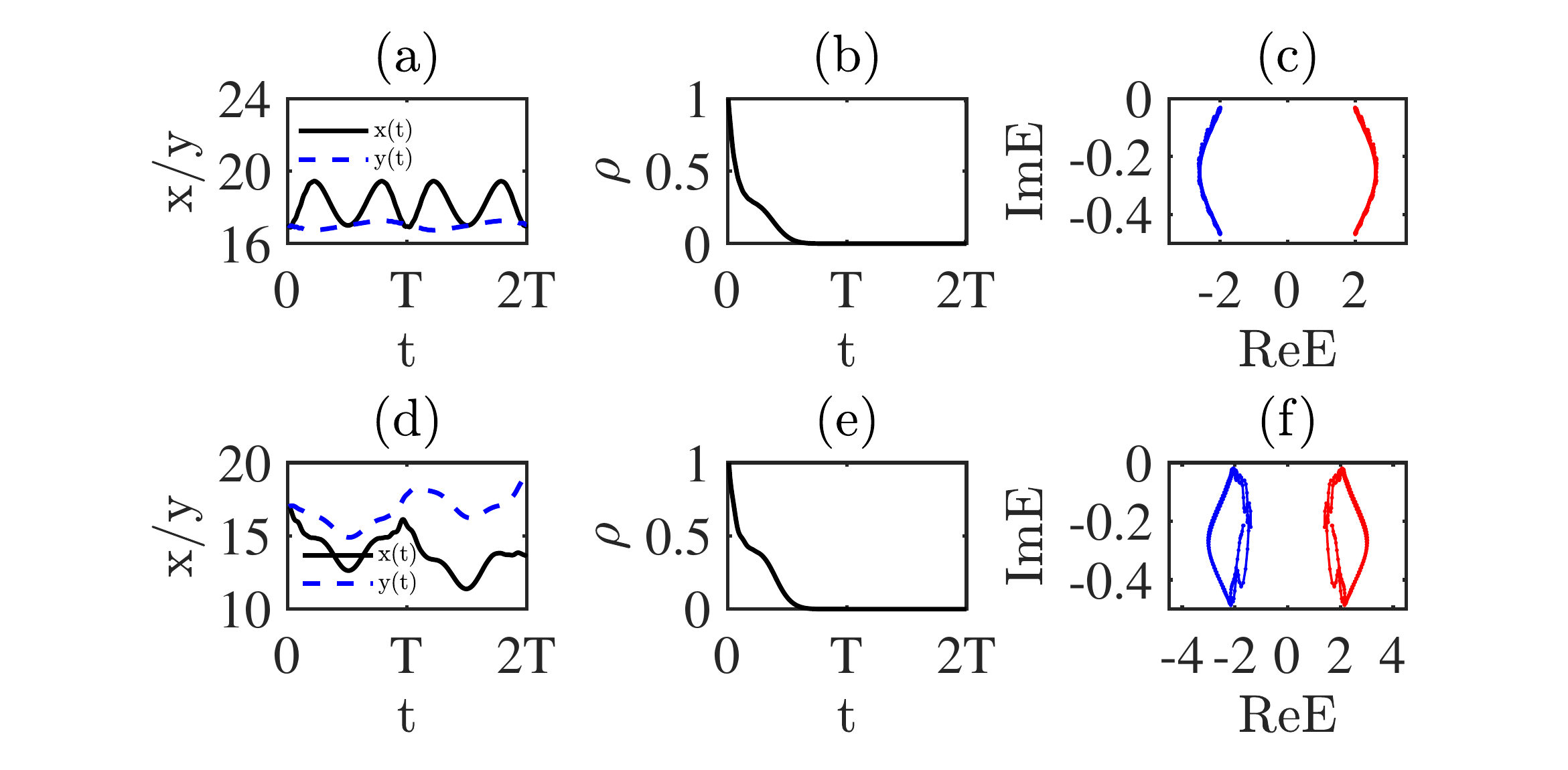}
 		\par\end{center}
 	\protect\caption{\label{fig:adiabatic}
\textbf{Non-Hermitian Bloch oscillations in the 2D $sp$ ladder lattice.} 	
(\textbf{a}) to (\textbf{c}) The center of mass, density and corresponding reconstructed energy trajectory for a static force $(F_x,F_y)=(0.25,0)$.
It is seen that the spatial profile of the wave-packet oscillates with a period of $T=2\pi/F_x$, yet its density vanishes after $t\geqslant T$.
(d) to (f) the same as (a) to (c), but with $(F_x,F_y)=(0.25,0.25)$.
Other parameters are the same as  Fig. 3 in the main text.  
} 
 \end{figure*}
 
An explicit condition for dynamically reconstructing the energy spectrum still requires more detailed inverstigation. Alternatively, we have numerically examined the wave-packet dynamics for different parameters, and found that an almost prefect reconstruction of the full spectrum can indeed be achieved, as shown in Fig.~\ref{fig:adiabatic} (and also Fig. 3 in the main text).
In Fig. \ref{fig:adiabatic}(a), a periodic oscillation is seen from the spatial profile of the wave-packet, justifying our definition of time period for non-Hermitian Bloch oscillations. The non-Hermitian loss still causes a significant decay of the wave-packet [Fig. \ref{fig:adiabatic}(b)], making it difficult for experimental observation. Nevertheless, full energy spectrum can still be extracted from the vanishing wave-packet, as shown in  in Fig. \ref{fig:adiabatic}(c).
In Fig. \ref{fig:adiabatic}(d) to (f), we illustrate another example where the energy trajectory forms some loops, corresponding to the presence of NHSE. The periodicity of the oscillations is less pronounced than in the previous case, yet a loop-like spectrum with nontrivial winding can still be obtained qualitatively.

\section{Symmetry breaking and transition of non-Hermitian skin effect}
 \begin{figure*}[t]
 	\begin{center}
 		\includegraphics[width=0.7\linewidth]{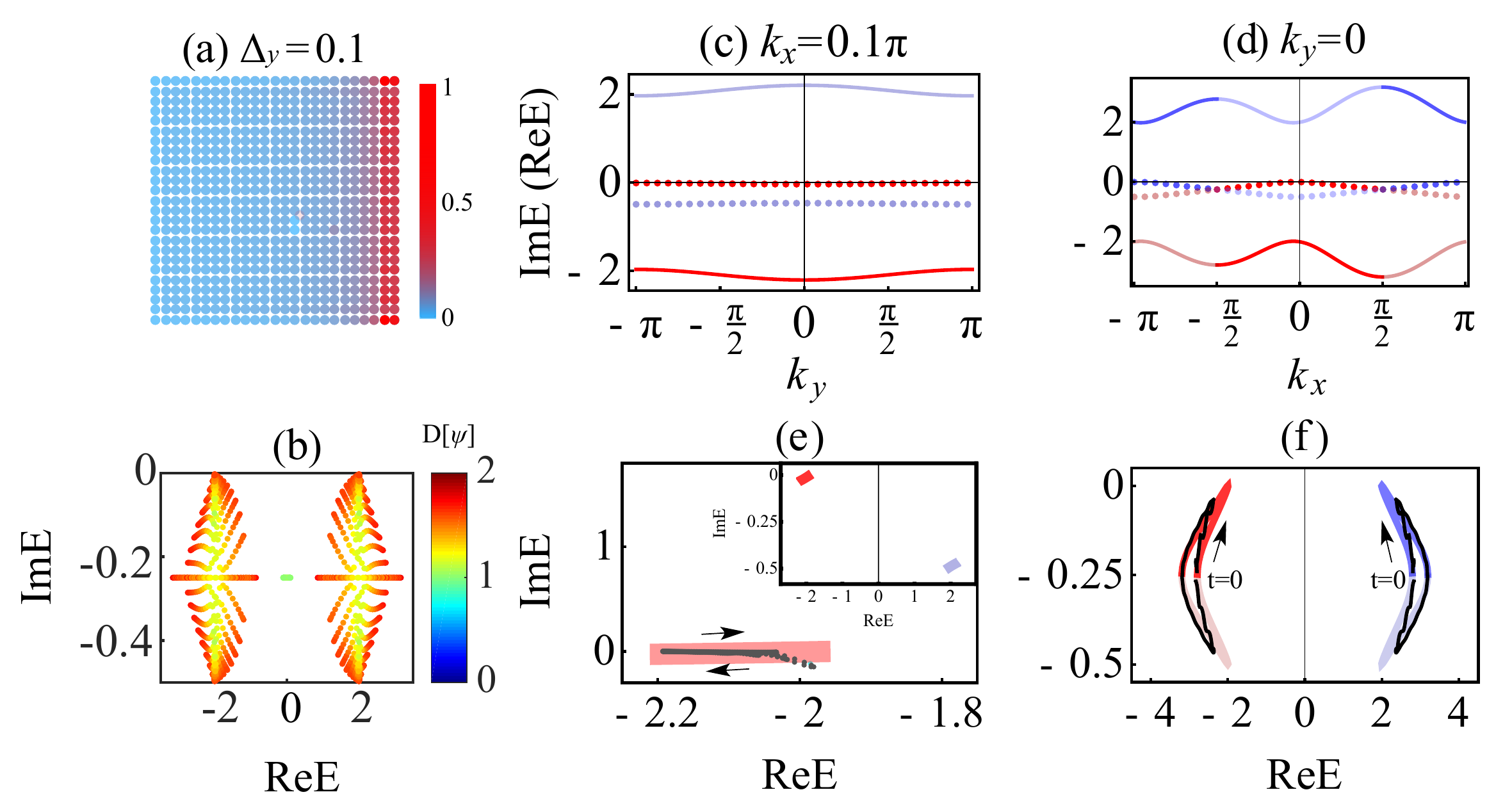}
 		\par\end{center}
 	\protect\caption{\label{fig:4} 
 		\textbf{The complex energies constructed from the Bloch oscillations along different directions for symmetry breaking phase.} 
		\textbf{Line NHSE, its corresponding spectra, and dynamical detection.}
		(\textbf{a}) The average density distribution and (\textbf{b}) fractional dimension of each eigenstate in the presence of line NHSE. 
		(\textbf{b}) fractional dimension of each eigenstate for $\Delta_y=0.1$. 
		(\textbf{c}) and (\textbf{d}) The real (lines) and imaginary (dots) band structures with $k_x=0.1\pi$ and $k_y=0$ respectively. Eigenvalues with larger (lower) imaginary part have greater (lesser) opacity.
		(e) and (f) The complex spectra corresponding to band structures with the same colors in (c) and (d), respectively,
		and energy trajectories obtained from the wave-packet dynamics (black curves).
		(e) displays only one of the two bands, whose full spectrum is shown in the inset.
		The initial wave packet are chosen to have $\bm{k}_0=(0.1\pi,0)$ in (e) and $(\pi/2,0)$ in (f), 
		with their corresponding static forces being $(F_x, F_y)=(0.3, 0)$ and $\pm(0, 0.3)$ respectively. 
		Plus (minus) sign reflects the evolution toward positive (negative) $k_x$ directions, reconstructing the energy band with blue (red)
		color.
		In each of these cases, $u_{\bm{k}_0}$ for the initial wave-packet is chosen to be the eigenstate of target energy band of $\mathcal{H}(\bm{k}_0)$. 
		Other parameters are the same as Fig.3 in the main text, with $\Delta_y=0.1$.} 
 \end{figure*}
In the Hermitian case with $\gamma=0$, the $sp$ ladder model exhibits a topological phase transition when rotating the atoms on individual sites, which induces tunneling between $s$ and $p_x$ orbitals within each unit cell and breaks the time-reversal symmetry and parity symmetry~\cite{Lixiao2013SM}.
The additional Hamiltonian term is given by
$
    \delta H = {{\rm{\Delta }}_y}\sum\limits_j {C_j^\dag } {\sigma _y}{C_j},
$
where $\Delta_y$ represents the tunneling strength. The total Hamiltonian becomes
\begin{equation}
  \mathcal{H}^{\prime}({k_x},{k_y}) = \mathcal{H}({k_x},{k_y})+\Delta_y \sigma_y
\end{equation}
with $\mathcal{H}({k_x},{k_y})$ given by Eq. 2 in the main text.
When $\gamma\neq 0$, this atom rotation also leads to a transition from GDSE to a line NHSE when $\Delta_y\neq0$.
That means the system changes from a generalized reciprocal skin effect class ($J=0$) to a non-reciprocal skin effect class ($J\neq 0$), where $J$ is the current function~\cite{Zhang2022sSM}. As shown in Fig. \ref{fig:4}(a), the average density of all eigenstates is now localized at the right boundary for square geometry.
The fractional dimension $D[\psi]$ of each eigenstate is demonstrated in Fig. \ref{fig:4}(b), 
where a majority of them has $1<D[\psi]\leqslant1.5$, further verifying the emergence of line NHSE.

Following the same strategy as in the main text, we consider 1D spectrum with fixed momentum in the other direction, as shown in Fig. \ref{fig:4}(c) and (d). 
Similar to Fig. 3 (a) and (b) in the main text,
the two bands are well separated in both real and imaginary energies in Fig. \ref{fig:4}(a), 
while have crossing points of imaginary energies in Fig. \ref{fig:4}(b).
However, as shown in Fig. \ref{fig:4}(d), the complex spectrum with fixed $k_y=0$ forms two loops with nontrivial winding in the complex energy plane, indicating the emergence of line NHSE along $x$ direction.
With similar choices of initial states and applied static forces, the absence and presence of nontrivial winding for spectra versus $k_y$ and $k_x$ can also be captured by the wave-packet dynamics, as shown in Fig. \ref{fig:4}(c) and (d), respectively. Thus the line NHSE induced by atom rotation in our model can also be probed from bulk dynamics.


\begin{thebibliography}{77}
\bibitem [{not()}]{note1}%
  \BibitemOpen
  \href@noop {} {\bibinfo  {journal} {Numerically, we find that small
  fluctuation of $\rho_b$ arises within within a certain parameter regime of
  theta, which may be associated with details of the lattice’s boundary
  structure and spectral properties for different theta, and is await further
  investigation}\ }\BibitemShut {NoStop}%
\end{thebibliography}
%

\end{widetext}
\end{document}